\documentclass[final,3p,times,11pt]{elsarticle}

\usepackage{amssymb}
\usepackage{amsmath}
\usepackage{amsmath,bm}
\usepackage{xcolor}
\usepackage{caption}
\usepackage{subcaption}
\usepackage{multirow}
\usepackage{graphicx,import}
\usepackage{siunitx}
\usepackage{float}
\usepackage{lipsum}
\usepackage{algorithm}
\usepackage{algpseudocode}
\usepackage{comment}
\usepackage{graphicx} 
\usepackage[version=4]{mhchem}
\usepackage[colorlinks=true, linkcolor=blue, citecolor=blue, urlcolor=blue]{hyperref}
\usepackage[capitalize]{cleveref}
\usepackage{epstopdf}
\usepackage{orcidlink}
\AppendGraphicsExtensions{.tif}

\setcitestyle{numbers}
\journal{Journal of Manufacturing Processes}

\begin{document}

\begin{frontmatter}



\title{Microstructure engineering of Ti-6Al-4V in laser powder bed fusion via 1D thermal modeling and supporting experiments}

\author[label1]{C. van der Linde\,\orcidlink{0000-0002-9385-2098}}
\author[label1]{I. Sideris\,\orcidlink{0000-0002-4584-9601}}
\author[label1]{L. Deillon\,\orcidlink{0000-0001-8374-4969}}
\author[label1]{M. Afrasiabi\corref{cor1}\,\orcidlink{0000-0003-1802-1857}}
\author[label1]{M. Bambach\,\orcidlink{0000-0002-8790-0807}}

\affiliation[label1]{organization={Advanced Manufacturing Lab, ETH Zurich},
            addressline={Leonhardstrasse 21}, 
            city={Zurich 8092},
            country={Switzerland}}




\begin{abstract}
The microstructure of Ti-6Al-4V has a decisive impact on its mechanical performance; however, controlling phase composition during Laser Powder Bed Fusion (LPBF) remains difficult because of the inherent localized and cyclic thermal history. To fully leverage the design flexibility of LPBF while maintaining an efficient process, it is desirable to tailor the microstructure directly through process-parameter optimization rather than relying on post-processing or in-situ heat treatments. Nevertheless, the large and multidimensional parameter space, combined with the limited availability of experimental data, makes this task particularly challenging. 
In this work, we develop an efficient computational framework that links process conditions to microstructure evolution by coupling a phase transformation model with a fast 1D finite-difference thermal model, enabling comprehensive insights into process-microstructure relations.
The framework predicts the fractions of $\alpha_s$, $\alpha_m$, and $\beta$ phases and is validated experimentally. A broad design of experiments covering 2,000 parameter combinations (spanning volumetric energy density, layer thickness, interlayer time, and build plate temperature) demonstrates how these parameters influence phase evolution and provides systematic practical guidelines for process design. The framework reproduces experimental trends with sufficient accuracy while being orders of magnitude faster than high-fidelity simulations, enabling rapid exploration of process–structure relationships in LPBF of Ti-6Al-4V.

\begin{figure}[H]
    \centering
    \includegraphics[width=\textwidth]{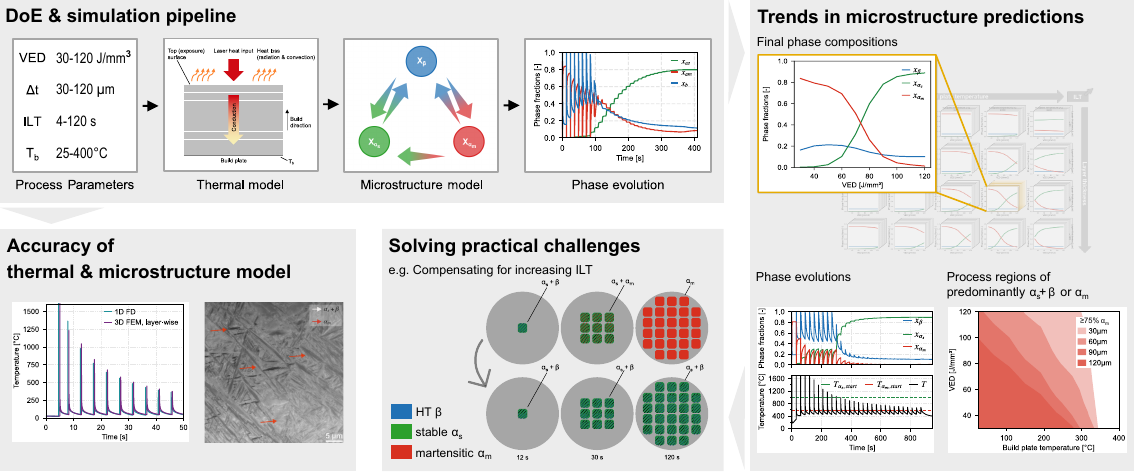}
\end{figure}

\end{abstract}



\end{frontmatter}


\section{Introduction}

Laser Powder Bed Fusion (LPBF) is one of the most promising Additive Manufacturing (AM) techniques for metallic materials, offering high design flexibility, near-net-shape fabrication, and excellent part quality. Due to its localized melting and rapid solidification, LPBF enables resource-efficient production of geometrically complex and customized parts, which has driven its adoption across multiple sectors, including aerospace, biomedical, and energy.  

Among the materials processed by LPBF, Ti-6Al-4V is the most widely used titanium alloy, valued for its combination of high specific strength, corrosion resistance, and biocompatibility \cite{esmaeilzadeh2023Insitu}. Ti-6Al-4V forms three main phases: equilibrium high-temperature $\beta$ phase, equilibrium low-temperature stable $\alpha_S$ as well as martensitic $\alpha^\prime$ or $\alpha_m$. The extreme cooling rates inherent to LPBF, as high as $10^6$~K/s, induce a predominantly martensitic $\alpha_m$ microstructure rather than the equilibrium duplex $\alpha_s+\beta$ microstructure. While this martensitic structure enhances tensile strength and fatigue strength \cite{munk2022Geometry}, it generally limits ductility. Consequently, numerous studies have investigated post-process heat treatments or in-situ control strategies to decompose $\alpha_m$ into a more ductile lamellar $\alpha_s+\beta$ microstructure \cite{esmaeilzadeh2023Insitu}.

Several studies found that LPBF microstructures are often spatially heterogeneous, strongly influenced by part geometry and thermal history. For instance, variations in cross-sectional area can shift the local microstructure from acicular martensite to a mixed $\alpha_s+\beta$ state due to local heat accumulation \cite{munk2022Geometry,nahr2023Geometrical,olleak2024Understandinga}. To mitigate such heterogeneities, process-control approaches such as adaptive power strategies \cite{khrenov2024Trajectory} and feedback-based temperature regulation \cite{nahr2025Advanced} have been developed, achieving more uniform phase distributions and mechanical responses. Beyond homogenization, emerging strategies aim to intentionally engineer spatially varying microstructures to leverage local phase control and thus local mechanical properties for functionally graded or architected components.

Microstructure tailoring in LPBF-produced Ti-6Al-4V has traditionally relied on post-processing approaches such as heat treatment \cite{vrancken2012Heat} or hot isostatic pressing (HIP) \cite{benzing2019Hot}. These methods effectively decompose the non-equilibrium $\alpha_m$ formed during rapid solidification into a lamellar $\alpha+\beta$ microstructure. However, post-process treatments add an additional process step and fail to exploit the design freedom LPBF offers in terms of local microstructure design, motivating strategies that enable similar phase transformations to be achieved during the LPBF process.

To address this, several studies have explored the potential of performing in-situ heat treatments during LPBF. Chen et al. \cite{chen2022Deciphering} introduced an additional reheating laser pass to locally control cooling rates, generating a layered structure consisting of $\alpha_s$, $\alpha_m$, and nano-sized $\beta$ precipitates. By adjusting interlayer times, these authors demonstrated that heat accumulation can promote the diffusive formation of lamellar $\alpha_s+\beta$ upon slow cooling from the $\beta$ phase. Similarly, Esmaeilzadeh et al. \cite{esmaeilzadeh2023Insitu} validated in-situ microstructural transitions through operando X-ray diffraction, confirming real-time $\alpha_m$ decomposition. Nonetheless, open questions remain regarding the spatial and temporal scales required to transition between distinct microstructural states. 

Alternatively, parameter-based strategies aim to induce desirable microstructures in the as-built state without in-situ heat treatment steps. Studies have shown that reducing scanning speed, increasing layer thickness, and adjusting focal offset distance or hatch spacing can locally decrease cooling rates and promote $\alpha_s+\beta$ formation \cite{haubrich2019Rolea,xu2015Additivea,barriobero-vila2017Inducing}. For instance, Xu et al. \cite{xu2015Additivea} achieved in-situ martensite decomposition through thick layers (\SI{60}{\micro\metre} and \SI{90}{\micro\metre}) and optimized energy input, attaining high strength and ductility. Dhiman et al. \cite{dhiman2024Microstructure} further demonstrated that high-power LPBF at \SI{600}{\watt} and a layer thickness of \SI{90}{\micro\metre} can yield  as-built lamellar microstructures. Yet, excessive heat accumulation, as observed by Medvedev et al. \cite{medvedev2025Interlayer}, may induce coarsening, impurity uptake, and embrittlement, highlighting the delicate balance between thermal management and manufacturability in process optimization.

Leveraging additional laser parameters offers another path for local microstructure tailoring. Zafari et al. \cite{zafari2019Controlling} demonstrated that high-energy pulsed laser modes can initiate $\alpha_m$ decomposition selectively within bulk Ti-6Al-4V. 
Moreover, Esmaeilzadeh et al. \cite{esmaeilzadehArchitected} showed that beam shaping can spatially modulate cooling rates in Ti-6Al-4V, enabling the creation of architected 3D microstructures with region-specific phase compositions.

While these experimental studies have advanced the understanding of LPBF-induced microstructures, their interpretability remains constrained. The complex and cyclic reheating behavior inherent to LPBF makes it difficult to relate post-mortem microstructure analysis to microstructure evolution. Hence, predictive microstructure models are crucial for bridging this gap. Furthermore, as in-process temperature fields are not directly observable in LPBF \cite{wood2022Controllability}, combining microstructure models with thermal models becomes indispensable for estimating internal states and guiding process control. Analytical thermal models, though limited by assumptions such as constant material properties and semi-infinite domains, serve as efficient tools for early parameter-space screening \cite{vanini2025Local}.

Microstructure modeling approaches for LPBF generally fall into statistical, phenomenological, and phase-field categories \cite{nitzler2021Novel,murgau2012Model}. Among these, phenomenological models offer a favorable compromise between accuracy and computational efficiency. Building upon extensive work in welding research \cite{kelly2004Thermala,fan2005Effect,charles2008Modelling,ahmed1998Phase,malinov2001Differential}, Murgau et al. \cite{murgau2012Model} introduced a comprehensive model for Ti-6Al-4V, combining diffusion-driven Johnson–Mehl–Avrami–Kolmogorov (JMAK) kinetics \cite{johnson1939reaction,avrami1941kinetics,kolmogorov1937statistical} with diffusionless Koistinen–Marburger transformations \cite{koistinen1959general}. This formulation accommodates arbitrary heating and cooling cycles, including partial and reversed transformations. Later refinements by Nitzler et al. \cite{nitzler2021Novel} reformulated the equations in transformation rate form, allowing temperature-dependent kinetic parameters and improving consistency with time–temperature–transformation diagrams.

Recent research has increasingly focused on coupling such models with part-scale thermal simulations to predict spatially resolved phase evolution. Examples include frameworks based on Finite Element Method (FEM) for LPBF \cite{yang2021Processstructure,promoppatum2022Understanding,lee2024Extreme}, Directed Energy Deposition \cite{babu2019Simulation}, Electron Beam Melting and Wire Arc Additive Manufacturing \cite{vastola2016Modelinga,charlesmurgau2019Temperature}. These studies successfully correlate predicted microstructures with experimental findings.
More recently, Proell et al. \cite{proell2024Highlya} demonstrated a scan-resolved, computationally efficient coupling of the microstructure model by Nitzler et al. \cite{nitzler2021Novel}  with LPBF thermal simulations. Beyond empirical kinetics, Noll et al. \cite{noll2024Thermodynamically} proposed a thermodynamically consistent formulation driven by dissipation functions, offering a generalizable framework for multiphase alloys such as Ti-6Al-4V.

While not focusing on Ti-6Al-4V, Mozaffar et al. \cite{mozaffar2023Differentiable} progress from merely analyzing to actively controlling microstructure evolution in a simplified case study. By employing a differentiable Finite Element Method (FEM) thermal model, they optimize the laser power so that as many material points as possible spend a target time span within a defined heat treatment temperature range.

As the studies mentioned above demonstrate, thermal modeling plays a central role in linking process parameters to microstructure evolution. Yet these studies applied computationally demanding thermal models, preventing a comprehensive process parameter study. In LPBF thermal modeling, approaches, such as the adaptive multiscale approach by Scheel et al. \cite{scheel2023Advancing} and the GPU-accelerated framework GO-MELT by Leonor et al. \cite{leonor2024GOMELTa}, have significantly improved simulation efficiency but still incur high computational costs. To address this, reduced-order 1D thermal models present efficient alternatives. These simplified 1D formulations are e.g. employed to guide dwell-time adjustments achieving reliable process control \cite{kavas2025Physicsaware} and mitigate distortion \cite{bierwischDevelopment}, while maintaining computational speed.

To summarize, most LPBF experimental studies provide only post-mortem microstructure information and, due to their high resource demand, yield sparse datasets that leave important questions about phase evolution unresolved. However, attempts to bridge this gap using microstructure models have so far relied on computationally intensive thermal simulations, limiting investigations to only a few parameter combinations and preventing the broad parameter sweeps needed to establish general microstructure design principles.

Based on these observations, we identified the following research questions:
\begin{enumerate}
    \item How accurately does the proposed 1D thermal–microstructure framework predict relevant thermal histories required to compute phase trends in Ti-6Al-4V LPBF parts with near-constant cross-sections?
    \item How do key processing parameters influence phase evolution and which combinations define the transition between predominantly $\alpha_m$ and predominantly $\alpha_s + \beta$ microstructures in the investigated process window?
    \item How can the developed framework be used to identify compensating parameter changes for practical LPBF microstructure-engineering tasks?
\end{enumerate}

To answer these questions, we develop an efficient computational framework for predicting and designing Ti-6Al-4V microstructures in LPBF by combining, to the best of the authors’ knowledge for the first time, a fast 1D finite-difference (FD) thermal model with a microstructure evolution model. The computational efficiency of this framework allows a design of experiments (DoE) of 2,000 process parameter combinations in contrast to the existing literature enabling a more systematic study. The DoE explores the effects of volumetric energy density (VED), layer thickness ($\Delta t$), interlayer time (ILT), and build plate temperature ($T_b$) on phase formation and evolution. The results are used to evaluate the model accuracies as well as to identify process–microstructure trends, and derive design guidelines based on this novel comprehensive approach with 2,000 samples. The framework is finally applied to a few practical LPBF scenarios.

The remainder of this manuscript is structured as follows. \cref{sec:meth} introduces the materials and methods, starting with the microstructure and thermal models and then outlining the Ti-6Al-4V microstructure study configuration and the LPBF print jobs used for validation. \cref{sec:results_discussion} presents and discusses the results, starting with the accuracy assessment of both the 1D thermal model and the microstructure predictions, followed by an analysis of trends observed across the process parameter space. Finally, the practical applicability of the framework is demonstrated through three representative case studies: (i) compensating for increasing interlayer times, (ii) varying microstructure with build height, and (iii) producing parts with distinct microstructures within a single build job.

 \section{Materials and methods \label{sec:meth}}

\subsection{Microstructure model \label{sec:microstructure_model}}
\begin{figure}[]
    \includegraphics[width=\textwidth]{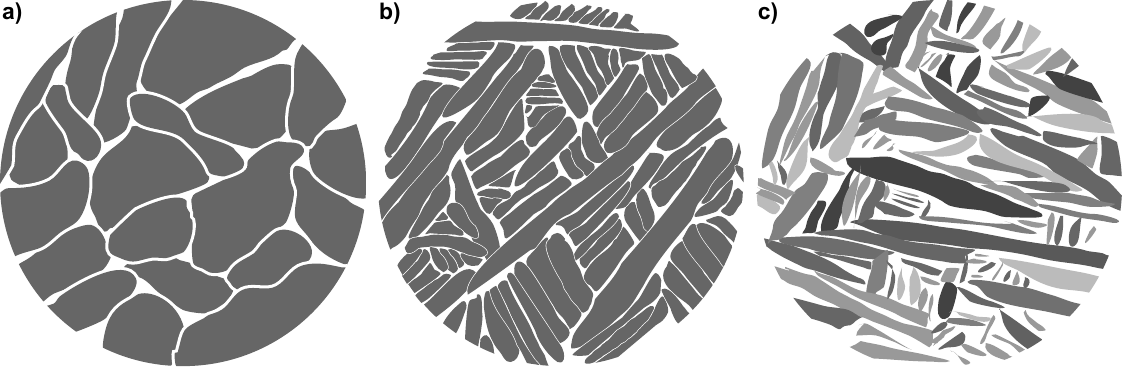}
    \caption{Schematic micrographs of (a) a globular $\alpha_s$ microstructure, (b) a lamellar $\alpha_s + \beta$ microstructure and (c) an $\alpha_m$ martensitic microstructure as observed in as-built LPBF processed Ti-6Al-4V.}
    \label{fig:microstructures}
\end{figure}

The used phenomenological microstructure model was originally proposed by Nitzler et al. \cite{nitzler2021Novel} and extended by Proell et al. \cite{proell2024Highlya}. This model predicts phase evolution in Ti-6Al-4V for a given temperature history. Three phases are distinguished: the high-temperature $\beta$ phase as well as two low-temperature $\alpha$-phases, namely stable $\alpha_{s}$ and martensitic/massive $\alpha_m$. These phases compose the microstructures of LPBF processed Ti-6Al-4V, which typically range from globular $\alpha$ microstructure (\cref{fig:microstructures}a), lamellar $\alpha + \beta$ microstructure (\cref{fig:microstructures}b) to $\alpha_m$ martensite microstructure (\cref{fig:microstructures}c). The key concepts of this model are summarized here. Further details are provided in \cite{nitzler2021Novel} and \cite{proell2024Highlya}.

\begin{figure*}[]
    \centering
    \includegraphics[width=\textwidth]{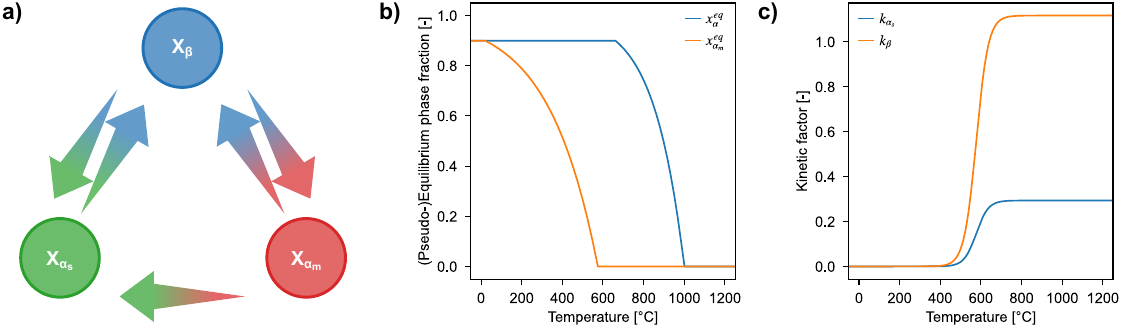}
    \caption{
    (a) Overview of Ti-6Al-4V microstructure model.
    (b) (Pseudo-)Equilibrium phase fractions $x_\alpha^{eq}$ and $x_{\alpha_m}^{eq}$ for accumulated $\alpha$ phase and $\alpha_m$ phase, respectively.
    (c) Diffusion rates $k_{\alpha_s}$ and $k_\beta$ as a function of temperature.
    }
    \label{fig:nitzler_model}
\end{figure*}

Upon solidification, the high temperature $\beta$ phase forms first, which then transforms either into the stable low temperature $\alpha_s$ phase under slow cooling rates 
or into the non-equilibrium $\alpha_m$ martensite under rapid cooling.
This simplified relationship is modeled in more detail in the used microstructure model, which accounts for cooling as well as reheating following any arbitrary time-temperature evolution. This requires modeling not only phase transformations from the high-temperature $\beta$ phase to the low-temperature $\alpha_s$ and $\alpha_m$ phases, but also the reverse transformations, resulting in the five phase transformations shown in \cref{fig:nitzler_model}a.

The kinetics of these phase transformations rely on the two temperature-dependent (pseudo-)equilibrium phase fractions $x_{\alpha_m}^{eq}$ and $x_\alpha^{eq}$. They are modeled based on the Koistinen-Marburger law and are displayed in \cref{fig:nitzler_model}b as a function of temperature. Generally, all phase fractions refer to the phase volume.

The phase transformations $\beta \rightarrow \alpha_s$, $\alpha_m \rightarrow \alpha_s$ and $\alpha_s \rightarrow \beta$ are diffusive transformations. They are modeled as modified logistic differential equations, here representatively shown for $\beta \rightarrow \alpha_s$ in \cref{eq:beta_to_alpha_s}: 

\begin{equation}
\label{eq:beta_to_alpha_s}
    \dot{x}_{\beta \rightarrow \alpha_s} = 
    \begin{cases} 
        k_{\alpha_s}(T) \left( x_{\alpha_s} \right)^{\frac{c_{\alpha_s}-1}{c_{\alpha_s}}} \left( x_\alpha^{\mathrm{eq}} - x_\alpha \right)^{\frac{c_{\alpha_s}+1}{c_{\alpha_s}}} & \text{if } x_\alpha < x_\alpha^{\mathrm{eq}} \\ 
        0 & \text{otherwise}
    \end{cases}
\end{equation}

$k_{\alpha_s}$ describes the diffusion rate as a function of temperature and is displayed in \cref{fig:nitzler_model}c (together with $k_\beta$ used for the inverse transformation). The experimentally determined factor $c_{\alpha_s}$ is 2.51. The second term relates to how much of the target phase, in this case $\alpha_s$, is already available, which corresponds to the available interface. The last term is the difference from the corresponding equilibrium phase fraction. 

Unlike diffusive phase transformations, the transformation $\beta \rightarrow \alpha_m$ and the opposite transformation $\alpha_m \rightarrow \beta$ are instantaneous phase transformations. The kinetic equations are described by Karush-Kuhn-Tucker conditions, here representatively shown for $\beta \rightarrow \alpha_m$: 

\begin{equation}
    \text{If } x_{\alpha_m} > 0: \quad 
    x_{\alpha_m} - x_{\alpha_m}^{\mathrm{eq}} \geq 0 \quad \land \quad 
    \dot{x}_{\beta \rightarrow \alpha_m} \geq 0 \quad \land \quad 
    (x_{\alpha_m} - x_{\alpha_m}^{\mathrm{eq}}) \cdot \dot{x}_{\beta \rightarrow \alpha_m} = 0
\end{equation}

\subsection{Thermal models \label{sec:thermal_methods}}
The majority of this study is based on a 1D FD thermal model with a layer-wise heat source, which is introduced in \cref{sec:1DFD_methods}. Further, a 3D FEM thermal model with a layer-wise heat source as well as a 3D FEM thermal model with a scan-resolved heat source, are included in this study for benchmarking (see \cref{sec:3DFEM_methods}).

\subsubsection{1D FD thermal model \label{sec:1DFD_methods}}
\begin{figure}[]
    \centering      
     \includegraphics[width=\linewidth]{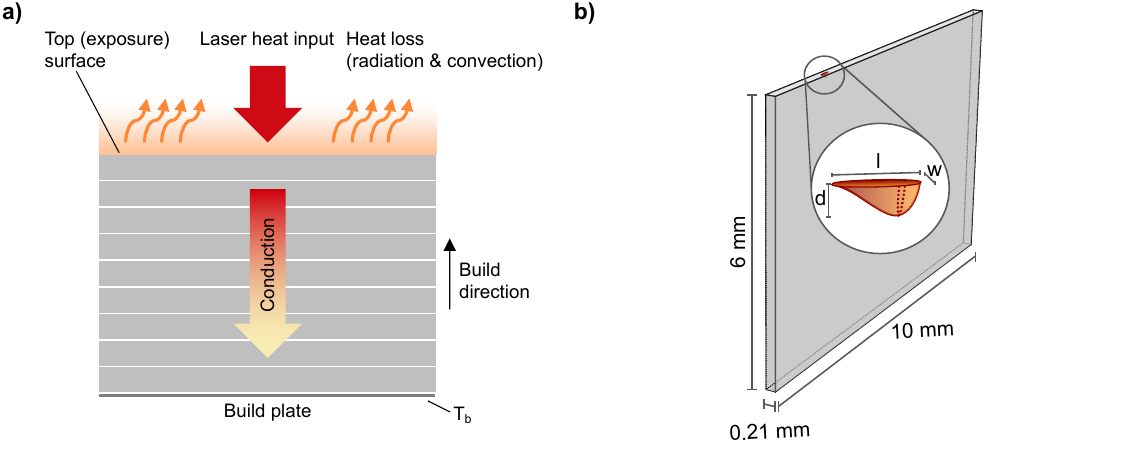}
    \caption{Set up of the 1D FD thermal model.}
    \label{fig:thermal_model_set_Up}
\end{figure}

The employed 1D FD thermal model is based on Fourier's law of heat conduction for the one-dimensional heat case, enabling fast layer-wise predictions at low computational cost. The model focuses on capturing heat flow in the z-direction, as illustrated schematically in \cref{fig:thermal_model_set_Up}a. This direction represents the dominant heat transport pathway in LPBF and governs the long-term thermal history of individual layers, which is the key information required to predict microstructure evolution throughout a build job. In contrast, lateral heat flow plays a comparatively minor role in determining global trends, since melt pool dimensions are orders of magnitude smaller than the characteristic layer and part dimensions. For example, experimental measurements in Ti-6Al-4V report average melt pool widths of about \SI{84}{\micro\metre} and depths of \SI{68}{\micro\metre} \cite{ransenigo2022meltpool}. In a high-energy single track setting, a melt depth of \SI{591}{\micro\metre} and a width of \SI{471}{\micro\metre}  was observed at a layer thickness of \SI{180}{\micro\metre}, laser power of \SI{230}{\watt} and a laser speed of \SI{55}{\milli\metre} \cite{brudler2024Systematic}.
\cref{fig:thermal_model_set_Up}b puts these characteristic scales into perspective.

To approximate the thermal conditions in an actual LPBF process, the boundary conditions are chosen as indicated in \cref{fig:thermal_model_set_Up}a: The heat input from the laser as well as a radiative and a convective boundary conditions are applied to the top surface (Neumann boundary conditions) and the build plate temperature is imposed at the bottom surface (Dirichlet boundary condition).
Assuming a temperature-independent thermal conductivity coefficient $k$ and a constant cross section $A$, the one-dimensional transient heat conduction equation including the Neumann boundary conditions can hence be expressed as:

\begin{equation}
    k \frac{\partial ^2 T}{\partial x ^2} 
    - \varepsilon \sigma_B (T^4 - T_{\infty}^4) \delta_s 
    - h (T - T_{\infty} ) \delta_s 
    + \dot q \delta_s 
    = \rho c_p \frac{\partial T}{\partial t} 
\end{equation}

where $T$ describes the temperature, $x$ the distance from the build plate, $t$ the time, $\rho$ the density and $c_p$ the heat capacity.
Further, $\sigma_B$ is the Stefan-Boltzmann constant, while $k$, $\varepsilon$ and $h$ are the material-dependent thermal conductivity coefficient, thermal emissivity coefficient and convective heat loss coefficient, respectively.  $T_{\infty}$ is the ambient temperature and $\dot q$ denotes the heat input from the laser. The delta function $\delta_s$ identifies the top surface, taking the value 1 at the top surface and 0 at all other locations. 

Discretizing this equation according to the FD method with a central-space differencing scheme and rearranging leads to:
\begin{align}
    \dot{q}_0 A 
    &+ hA(T_{\infty} - T_m^i)\delta_S 
    + \varepsilon \sigma A({T_{\infty}}^4 - {T^i_m}^4) \delta_S \notag \\
    &+ kA \frac{T_{m-1}^i - T_m^i}{\Delta x} 
    + kA \frac{T_{m+1}^i - T_m^i}{\Delta x}
    = \rho A \Delta x c_p \frac{T_m^{i+1} - T_m^i}{\Delta t}
\end{align}

To eventually enable diagonalization of the system, the radiation term is linearized by a Taylor expansion:  

\begin{equation}
    Q_{rad} = \varepsilon \sigma A({T_{\infty}}^4 + 3T_{ref}^4 - 4T_{ref}{T^i_m}^3) \delta_S 
\end{equation}

where $T_{ref}$ is a reference temperature that was set to the build plate temperature. With this adjustment, the system can now be written as:

\begin{equation}
    \dot \theta = L \theta + Q
    \label{eq:exp_matrix_form}
\end{equation}
where $\theta$ is a state vector of size N containing the temperatures of the N elements of the system. $L$ is the system matrix of size NxN that maps the inherent temperature dynamics of the system. $Q$ contains the temperature-independent parts of the heat input as well as radiative and convective terms and is a vector of size N. 

Via eigenvalue decomposition of $A$ to $A = Z \Lambda Z^{-1}$, this system of differential equations can be transformed into a decoupled first order ordinary differential equation system:

\begin{equation}
    \dot v = \Lambda v + Z^{-1}Q
    \label{eq:decoupled_matrix_form}
\end{equation}

where $\Lambda$ is the eigenvalue matrix of size NxN and $Z$ is a NxN matrix whose columns contain the eigenvectors of $A$. This transformed system can be solved analytically. Treating $Q$ as a Dirac delta function $\delta (t)$ that gives an instantaneous heat impulse to the system similar to a laser leads to: 

\begin{equation}
    v(t)= (v_0 + Q) \cdot e^{\lambda t}
\end{equation}

\subsubsection{3D FEM thermal models \label{sec:3DFEM_methods}}
To validate the 1D FD thermal model described above, it was compared to two FEM based thermal models for LPBF, both implemented in the commercial software Abaqus. 

The first model is a scan-resolved thermal model developed by Scheel et al. \cite{scheel2023Advancing}, who provide furhter details. This model uses an adaptive-local/global multiscale modeling approach that resolves scan tracks with a mesh size of \SI{10}{\micro\metre} × \SI{50}{\micro\metre} × \SI{30}{\micro\metre} and a Goldak heat source. The model was experimentally calibrated via in-situ thermocouple temperature measurements for a thin wall made of Hastelloy X. Among the three models in the present study, this thermal model is associated with the highest accuracy as well as highest computational cost, making it approximately 10\textsuperscript{3}-10\textsuperscript{4}x slower than the 1D FD model. 

Further, a 3D FEM thermal model was developed for the present study that uses a layer-wise heat source. The layer-wise heat source releases the total energy that the laser inputs into the part over an impulse length of \SI{0.001}{s}.
The build plate temperature is modeled as a Dirichlet boundary condition at the bottom surface. The outer surfaces of the part exhibit convection and radiation. The model uses temperature-independent material properties. The mesh size of the domain exposed by the laser is at most \SI{500}{\micro\metre} × \SI{500}{\micro\metre} × \SI{30}{\micro\metre}. 

\subsubsection{Considered geometries and materials}
\begin{figure*}[]
    \centering
    \includegraphics[width=\textwidth]{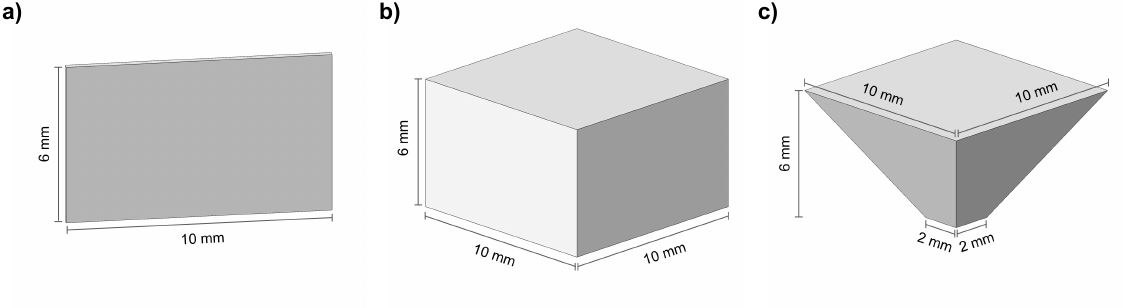}
    \caption{Part geometries investigated to compare three considered thermal models:  (a) thin wall, (b) cuboid, (c) inverted pyramid.}
    \label{fig:geometry_thermal_model}
\end{figure*}

Three different part geometries are considered to compare the predictions of the 1D FD thermal model to the two 3D FEM thermal models mentioned above. The selected geometries are characterized by distinct thermal conditions and are representative of many  part and feature geometries in LPBF. The considered geometries are:

\begin{enumerate}
    \item a thin wall with dimensions \SI{10}{mm} x \SI{0.21}{mm} x \SI{6.3}{mm} (see \cref{fig:geometry_thermal_model}a) in accordance with the study from Scheel et al. \cite{scheel2023Advancing},
    \item a cuboid with dimensions \SI{10}{mm} x \SI{10}{mm} x \SI{6.3}{mm} (see \cref{fig:geometry_thermal_model}b),
    \item an inverted pyramid with a \SI{2}{mm} x \SI{2}{mm} bottom surface that expands to \SI{10}{mm} x \SI{10}{mm} at the top surface over a height of \SI{6.3}{mm} (see \cref{fig:geometry_thermal_model}c).
\end{enumerate}

The material properties used for all three models are summarized in  \cref{tab:material_properties_HX_Ti64}. 
Since the scan-resolved 3D FEM thermal model was calibrated for Hastelloy X, the corresponding material parameters were included in addition to Ti-6Al-4V. 
The thermal conductivity was determined according to the temperature-dependent predictions for Hastelloy X and according to the most relevant temperature range of $\beta$ tranisitioning to $\alpha_s$ for Ti-6Al-4V.

\begin{table*}[h!]
\caption{Model parameters used for Ti-6Al-4V as well as Hastelloy X in all three employed thermal models. *The scan-resolved 3D FEM model uses a temperature-dependent thermal conductivity.}
\label{tab:material_properties_HX_Ti64}
\centering
\begin{tabular}{llrrcrr}
\hline
\bf Property               & \bf Unit                         & \multicolumn{2}{c}{\bf Ti-6Al-4V}              &  & \multicolumn{2}{c}{\bf Hastelloy X}        \\ \hline
Density                & \si{\kilogram\per\meter\cubed}        & 4090 & \cite{proell2024Highlya}        &  & 8220 & \cite{scheel2023Advancing}      \\
Specific heat capacity & \si{\joule\per\kilogram\per\kelvin}   & 1130 & \cite{proell2024Highlya}        &  & 486  & \cite{scheel2023Advancing}      \\
Conv. heat transfer coefficient       & \si{\watt\per\meter\squared\per\kelvin} & 20   & \cite{olleak2024Understandinga} &  & 25   & \cite{scheel2023Advancing}      \\
Absorption coefficient & \si{\percent}                         & 60   & \cite{olleak2024Understandinga} &  & 48   & \cite{scheel2023Advancing}      \\
Thermal conductivity   & \si{\watt\per\meter\per\kelvin}       & 20   &                                   &  & 15*  &                                   \\ \hline
\end{tabular}
\end{table*}

\subsection{Parameter study design for Ti–6Al–4V microstructure simulations \label{sec:forward_study_methods}}
\begin{figure*}[]
    \centering
    \includegraphics[width=\linewidth]{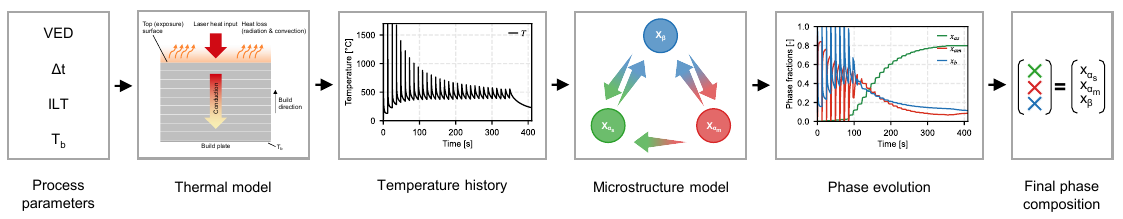}
    \caption{Pipeline of microstructure design framework.}
    \label{fig:pipeline_framework}
\end{figure*}

To study the phase evolution as well as final phase composition in Ti-6Al-4V, the 1D FD thermal model described in \cref{sec:1DFD_methods} and the microstructure model described in  \cref{sec:microstructure_model} are combined as illustrated in the pipeline depicted in  \cref{fig:pipeline_framework}. The corresponding code is available on \href{https://github.com/C-vdL/microstructure_engineering_lpbf_Ti6Al4V}{github}.

The four studied LPBF process parameters VED, $T_b$, $\Delta t$ and ILT are fed into the thermal model. The thermal model predicts the temperature evolution the printed part experiences throughout the build job. Subsequently, that temperature history is used as an input for the corresponding microstructure model, which predicts the evolution of phase fractions throughout the build job. From that evolution, the final phase fractions can be extracted. 

This pipeline was used to conduct a full-factorial design of experiment with the four considered LPBF process parameters listed above and the levels stated in \cref{tab:doe_forward_study}. This design of experiments comprises 2,000 process parameter combinations, which is equivalent to 2,000 printed parts that need to be distributed over at least 200 build jobs. 

\begin{table*}[t]
\centering
\caption{Parameters and corresponding levels considered in the full-factorial design of experiment of the conducted Ti-6Al-4V microstructure study.}
\label{tab:doe_forward_study}
\begin{tabular}{l l l r r r r r r r r r r}
\hline
\bf Parameter & \bf Abbr. & \bf Unit & \multicolumn{10}{l}{\bf Levels} \\
\hline
Volumetric energy density & VED & \si{\joule\per\milli\meter\cubed} & 30 & 40 & 50 & 60 & 70 & 80 & 90 & 100 & 110 & 120 \\
Build plate temperature & $T_b$ & \si{\celsius} & 25 & 100 & 200 & 300 & 400 &   &   &   &   &   \\
Layer thickness & $\Delta t$ & \si{\micro\meter} & 30 & 60 & 90 & 120 &   &   &   &   &   &   \\
Inter-layer time & ILT & \si{\second} & 4 & 6 & 8 & 10 & 12 & 20 & 30 & 60 & 90 & 120 \\
\hline
\end{tabular}
\end{table*}

The study investigates a part with a total height of \SI{18.9}{mm} and analyzes the phase composition at a representative layer of the bulk, located at a height of \SI{10.8}{\mm}. 

\subsection{LPBF print job \label{sec:experimental}}
\begin{figure}[]
    \centering
    \includegraphics[width=0.5\linewidth]{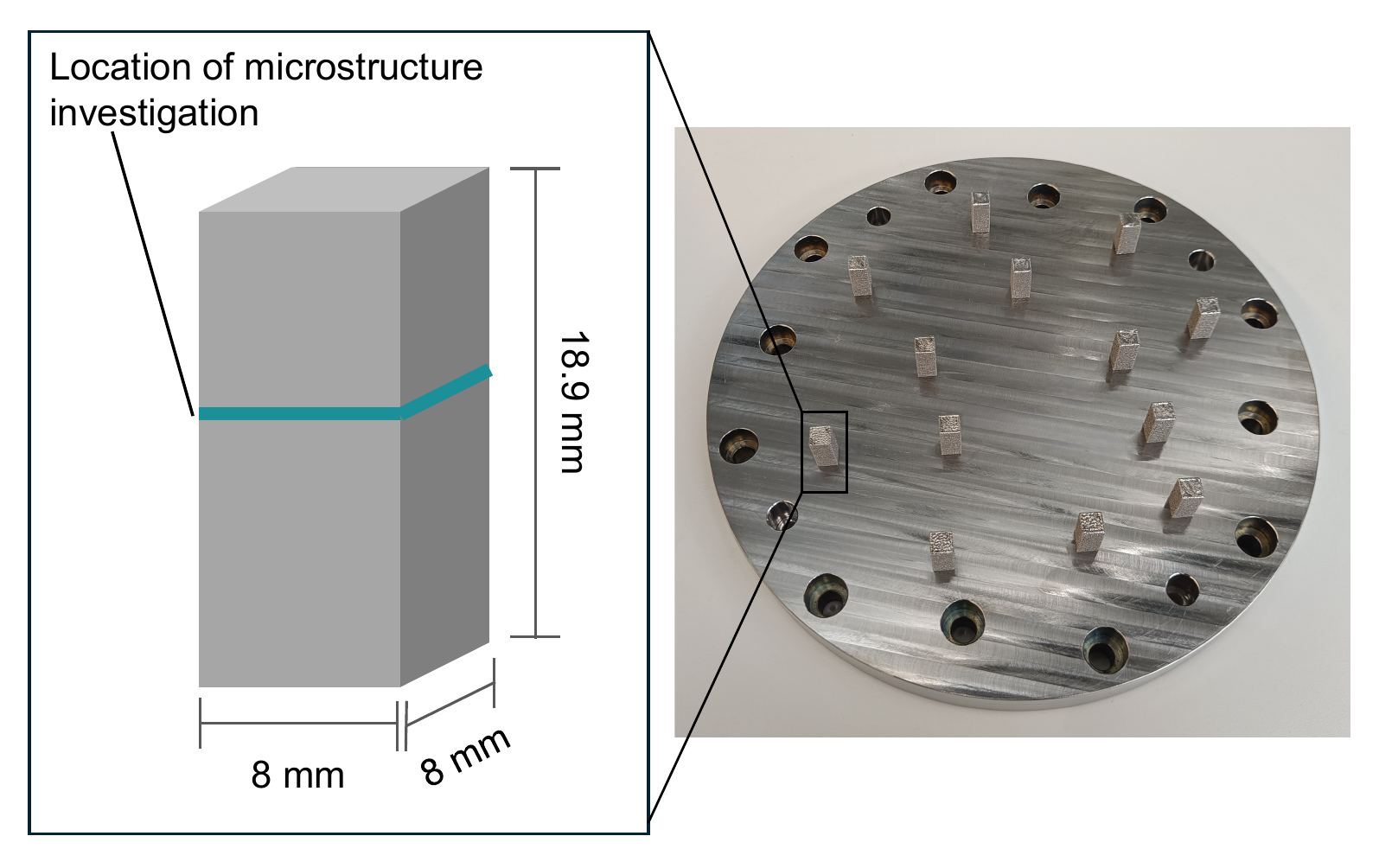}
    \caption{Geometry and dimensions of printed parts.}
    \label{fig:part_geometry}
\end{figure}

Ti-6Al-4V powder (thyssenkrupp Materials AG, Switzerland) with particle size \SI{63}{\micro\metre} $\pm$ \SI{20}{\micro\metre} was used in this study to fabricate cuboidal samples (\SI{8}{mm} x \SI{8}{mm} x \SI{18.9}{mm}) (see  \cref{fig:part_geometry}). 

The samples were fabricated using an Aconity Midi+ LPBF machine (Aconity3D GmbH, Germany) with a continuous-wave Gaussian-mode fiber laser and a wavelength of \SI{1080}{\nano\metre} (nLIGHT Alta, USA). All samples were produced with the same printing parameters listed in \cref{tab:exp_fixed_parameters}, except the laser speed.
\begin{table}
\centering
\caption{Process parameter employed in the two conducted build jobs.}
\label{tab:exp_fixed_parameters}
\begin{tabular}{l c r r}
\hline
\bf Process Parameter & \bf Unit & \bf Build job 1 & \bf Build job 2 \\
\hline
ILT & \si{\second} & 15 & 30 \\
Build plate temperature & \si{\celsius} & 200 & 200 \\
Laser power & \si{\watt} & 485 & 485 \\
Layer thickness & \si{\micro\meter} & 90 & 90 \\
Hatch spacing & \si{\micro\meter} & 75 & 75 \\
\hline
\end{tabular}
\end{table}
It is worth noting that the laser power was set to \SI{485}{\watt}, which is the maximum of the used laser. Further, the layer thickness was set to \SI{90}{\micro\metre}. Both parameter choices were based on the study by Dhiman et al. \cite{dhiman2024Microstructure}, which showed how the high power LPBF regime with a laser power of \SI{600}{\watt} and a layer thickness of \SI{90}{\micro\meter} enables the formation of lamellar $\alpha_s+\beta$ as-built microstructures with suitable dimensional accuracy. 

The laser speed, the only not fixed process parameter, was varied from \SI{800}{mm/s} (build job 1) and \SI{600}{mm/s} (build job 2) to \SI{1800}{mm/s} in steps of \SI{100}{mm/s}, leading to VEDs between \SI{39.92}{\joule\per\milli\meter\cubed} and \SI{119.25}{\joule\per\milli\meter\cubed} (see  \cref{tab:exp_laser_speeds_VED}). 
\begin{table*}[t]
\centering
\caption{Laser speed and resulting VED of samples in conducted build jobs.}
\label{tab:exp_laser_speeds_VED}
\resizebox{\textwidth}{!}{
\begin{tabular}{l c c c c c c c c c c c c c c}
\hline
\bf Laser speed & [mm/s] & 600 & 700 & 800 & 900 & 1000 & 1100 & 1200 & 1300 & 1400 & 1500 & 1600 & 1700 & 1800 \\
\hline
\bf VED & [J/mm$^3$] & 119.75 & 102.65 & 89.81 & 79.84 & 71.85 & 65.32 & 59.88 & 55.27 & 51.32 & 47.90 & 44.91 & 42.27 & 39.92 \\
\hline
\end{tabular}
}
\end{table*}
To minimize the influence of fume accumulation during fabrication, samples produced with comparatively high VEDs were positioned on the argon gas outflow side of the build chamber. The argon inflow was regulated to maintain the oxygen concentration below \SI{0.1}{wt.\percent} throughout the process. A meander scanning strategy with an interlayer rotation of \SI{67}{\degree} was applied.
To further study the effect of the interlayer time on microstructural evolution, two build jobs with different interlayer time of (1) \SI{15}{\second} and (2) \SI{30}{\second} were conducted.

After fabrication, the samples were removed from the build plate by electrical discharge machining, vertically cut in half and hot mounted. To prepare the samples for microstructure characterization, they were ground using 120 and 500 grit SiC paper and subsequently polished with a diamond suspension of \SI{9}{\micro\metre} and a solution of colloidal silica (\SI{0.25}{\um}) and \SI{30}{vol\percent} \ce{H2O2} (volume ratio 4:1).  
The microstructure was imaged using a Zeiss EVO 10 Scanning Electron Microscope (SEM) in backscattered electron (BSE) mode. Like in the full-factorial microstructure study, the microstructure is investigated at a build height of \SI{10.8}{mm} as indicated in \cref{fig:part_geometry}, which equals layer 120 in the conducted 210 layer build jobs. 

\section{Results and discussion \label{sec:results_discussion}}

\subsection{Accuracy of fast 1D FD thermal model \label{sec:accuracy_thermal}}

To assess the trade-off between computational efficiency and thermal accuracy with regard to microstructure modeling, three thermal models for LPBF were compared, namely (i) a 3D FEM model with a scan-resolved heat source, (ii) a 3D FEM model with a layer-wise heat source, and (iii) a simplified 1D FD model with a layer-wise heat source.

\begin{figure}[]
    \centering
    \includegraphics[width=\textwidth]{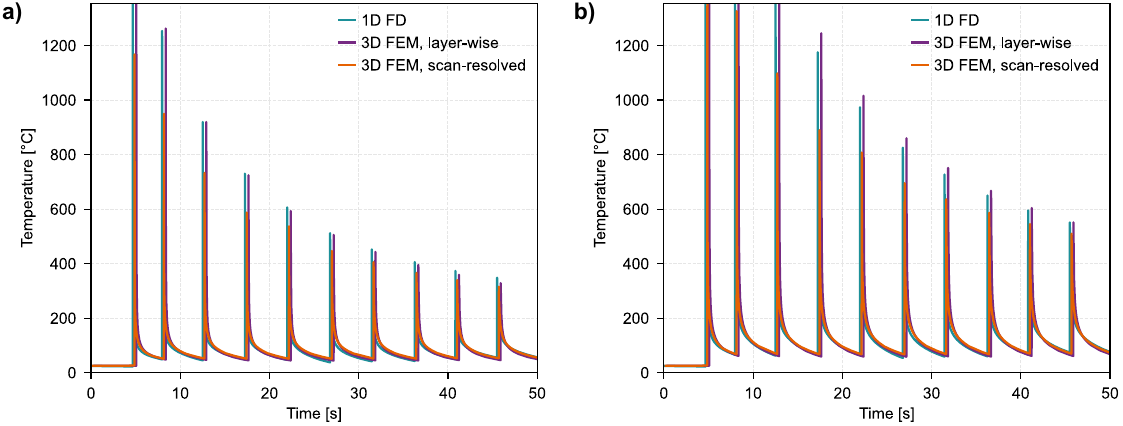}
    \caption{Thermal simulation results for Hastelloy-X thin wall geometry at layer 201 applying (a) \SI{52}{\joule\per\milli\meter\cubed} and (b) \SI{86}{\joule\per\milli\meter\cubed}.}
    \label{fig:thermal_sims}
\end{figure}

Since the multi-scale approach of the scan-resolved model by Scheel et al. \cite{scheel2023Advancing} was tailored for a thin wall geometry and calibrated for Hastelloy~X, thermal predictions from this model are only incorporated for that geometry and alloy. 
\cref{fig:thermal_sims} presents the temperature histories of layer 201 during the exposure and deposition of layer 201 to 210, for two representative volumetric energy densities \SI{53}{\joule\per\milli\meter\cubed} (\cref{fig:thermal_sims}a) and \SI{86}{\joule\per\milli\meter\cubed} (\cref{fig:thermal_sims}b). All temperature profiles are capped at the liquidus temperature of Hastelloy~X at \SI{1355}{\celsius}, as higher temperatures are physically irrelevant for solid-state process prediction.

When comparing the 1D FD layer-wise heat source model and the 3D layer-wise heat source model, both the peak temperatures and the cool-down phases show a high degree of agreement. The cooling profiles are nearly congruent with a Mean Average Percentage Error (MAPE, based on temperatures in K and assuming the respective highest fidelity model to be ground truth) of \SI{1.2}{\percent} (\SI{53}{\joule\per\milli\meter\cubed}) and \SI{1.4}{\percent} (\SI{86}{\joule\per\milli\meter\cubed}). The peak temperatures show minor deviations with Average Percentage Errors (APE, based on temperatures in K and assuming the respective highest fidelity model to be ground truth) ranging from \SI{-0.04}{\percent} to \SI{3.30}{\percent}. 

When now comparing the two layer-wise heat source models to the scan-resolved 3D FEM simulations, the cooling phase remains highly consistent with MAPEs of \SI{1.4}{\percent} (\SI{53}{\joule\per\milli\meter\cubed}) and \SI{1.0}{\percent} (\SI{86}{\joule\per\milli\meter\cubed}). However, the peak temperatures predicted by the scan-resolved model are lower. For instance, the APEs of the third peak, which is below the liquidus temperature for both considered VED, are \SI{18.51}{\percent} (\SI{53}{\joule\per\milli\meter\cubed}) and \SI{27.78}{\percent} (\SI{86}{\joule\per\milli\meter\cubed}). These values decrease throughout the build with APEs of \SI{5.57}{\percent} (\SI{53}{\joule\per\milli\meter\cubed}) and \SI{5.25}{\percent} (\SI{86}{\joule\per\milli\meter\cubed}) for the tenth peak. Note that these temperature differences only occur for fractions of a second around the peak temperature and continue to decrease during the build process beyond the low number of ten layers considered here. The deviations are attributed to lateral heat dissipation along the melt track, which cannot be represented in layer-wise heat source models where heat input is spatially averaged and happens simultaneously over the entire top surface. Further, the maximum temperature per layer will be overestimated as the total exposure time is reduced in layer-wise heat source models, which was pointed out by Nahr et al. \cite{nahr2023Geometrical}. These effects are amplified by the box size of \SI{100}{\micro\metre} used to average the nodal temperatures in the thermal models.

\begin{figure}[]
    \centering
    \includegraphics[width=\textwidth]{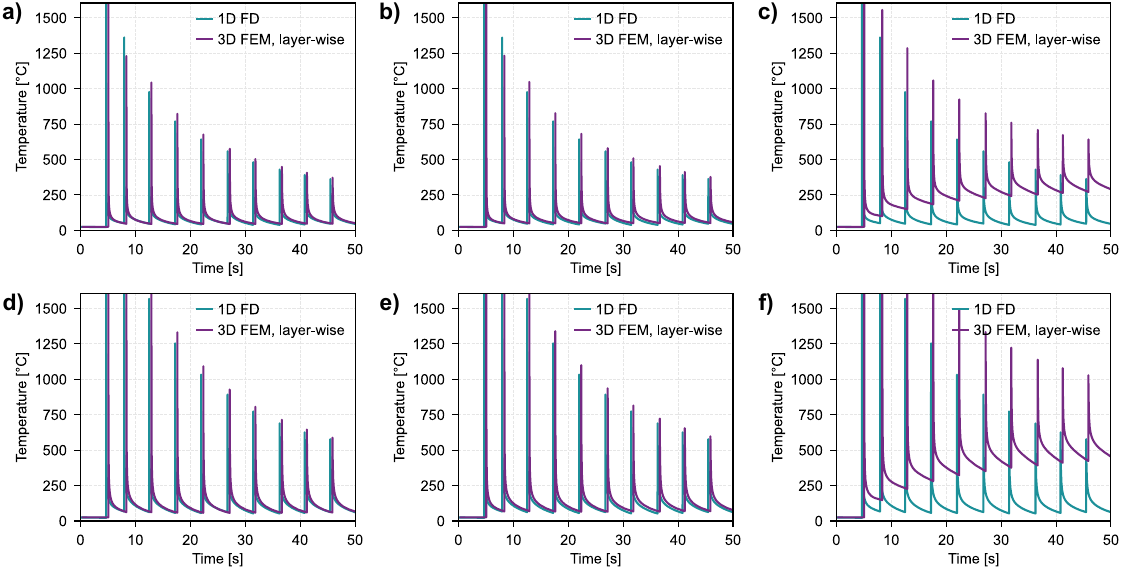}
    \caption{Thermal simulation results for different geometries and parameter sets:  
    (top row) \SI{52}{\joule\per\milli\meter\cubed},  
    (bottom row) \SI{86}{\joule\per\milli\meter\cubed};  
    geometries: (a,d) thin wall, (b,e) cuboid, (e,f) inverted pyramid.}
    \label{fig:thermal_grid}
\end{figure}

Extending the considered geometries, \cref{fig:thermal_grid} compares the predicted thermal histories of all three part geometries: the thin wall (see \cref{fig:thermal_grid}, left), the cuboid (see \cref{fig:thermal_grid}, middle), and the inverted pyramid (see \cref{fig:thermal_grid}, right). As explained above, this comparison is limited to the 1D FD and 3D FEM models with a layer-wise heat source. Again, VED values of \SI{53}{\joule\per\milli\meter\cubed} (see \cref{fig:thermal_grid}a, \cref{fig:thermal_grid}b, \cref{fig:thermal_grid}c) and \SI{86}{\joule\per\milli\meter\cubed} (see \cref{fig:thermal_grid}d, \cref{fig:thermal_grid}e, \cref{fig:thermal_grid}f) are evaluated.
Note that differences in temperature above the liquidus temperature, and even exact temperature values above the $\beta$-transus temperature of \SI{1000}{\celsius}, are insignificant for solid-state phase evolution in Ti-6Al-4V. 

For the thin wall as well as cuboid geometries, the peak temperatures and cooling curves are nearly congruent within the relevant range. The APE ranges from \SI{-5.26}{\percent} to \SI{-1.26}{\percent} for the peak temperatures, while MAPE ranges from \SI{0.7}{\percent} to \SI{2.0}{\percent} for the cooling periods. This confirms that for geometries with fixed cross-sections and limited thermal gradients, the simplified 1D formulation is sufficiently accurate to capture the relevant thermal behavior.

In contrast, for the inverted pyramid geometry, where the cross-section increases continuously with build height, only the 3D FEM model reproduces the progressive heat accumulation toward the upper layers among the two considered thermal models. The 1D FD model, lacking spatial resolution in the lateral directions, fails to represent this effect. Therefore, it predicts significantly lower temperatures throughout the build reaching an MAPE as high as \SI{36.6}{\percent} for \SI{86}{\joule\per\milli\meter\cubed} for the cool down phase and an APE of e.g. \SI{-30.52}{\percent} for the tenth peak at \SI{53}{\joule\per\milli\meter\cubed}.

From a computational standpoint, the differences in model complexity translate into substantial efficiency gains. The 1D FD model is approximately $10^3$~times faster than the 3D FEM model with the  layer-wise heat source, and roughly $3\times10^4$~times faster than the scan-resolved 3D model, when comparing the deposition of ten layers. 
To put these efficiency gains into perspective, a look at the heat transfer dynamics in LPBF is needed: 
LPBF pool dimensions are very small (on the scale of \SI{100}{\micro\meter}'s) in contrast to part dimensions or even layer areas. Thus, the lateral heat flow from the melt pool is insignificant compared to the heat flow in z-direction. This heat flow in z-direction governs the phase evolution throughout the build job, since it determines the temperature history of each layer in the part. Both the 3D and 1D thermal model with layer-wise heat sources can predict this. However, they do rely on VED instead of the individual process parameters that VED consists of and can hence not address their individual influences. 
This limitation is naturally offset by the significant speed advantage offered by VED-based analyses. In practice, it is rarely prohibitive, as experimental studies often deliberately use VED to simplify the parameter space while still capturing the key effects of process parameters for many applications.

Further, this accuracy assessment demonstrates that a 3D model is required to predict heat accumulation and geometric effects in non-uniform structures such as pyramids or lattice features. Nevertheless, the 1D FD model provides good accuracy for geometries with constant cross-sections and moderate complexity, where it can capture heat flow in z-direction similarly well as a 3D model.
Hence, despite its simplicity, the 1D FD model remains sufficient and fit-for-purpose for the present process parameter study focusing on these constant cross-section geometries, offering a practical balance between accuracy and computational cost for predictive microstructure simulations in LPBF. Similarly, Lee et al. \cite{lee2024Extreme} had good agreement between their moving heat source and line heat source in terms of temperature profiles and resulting phase evolution predictions for LPBF of Ti-6Al-4V. 

\subsection{Accuracy of microstructure predictions \label{sec:accuracy_microstructure}}
\begin{figure}[]
    \centering
    \includegraphics[width=\textwidth]{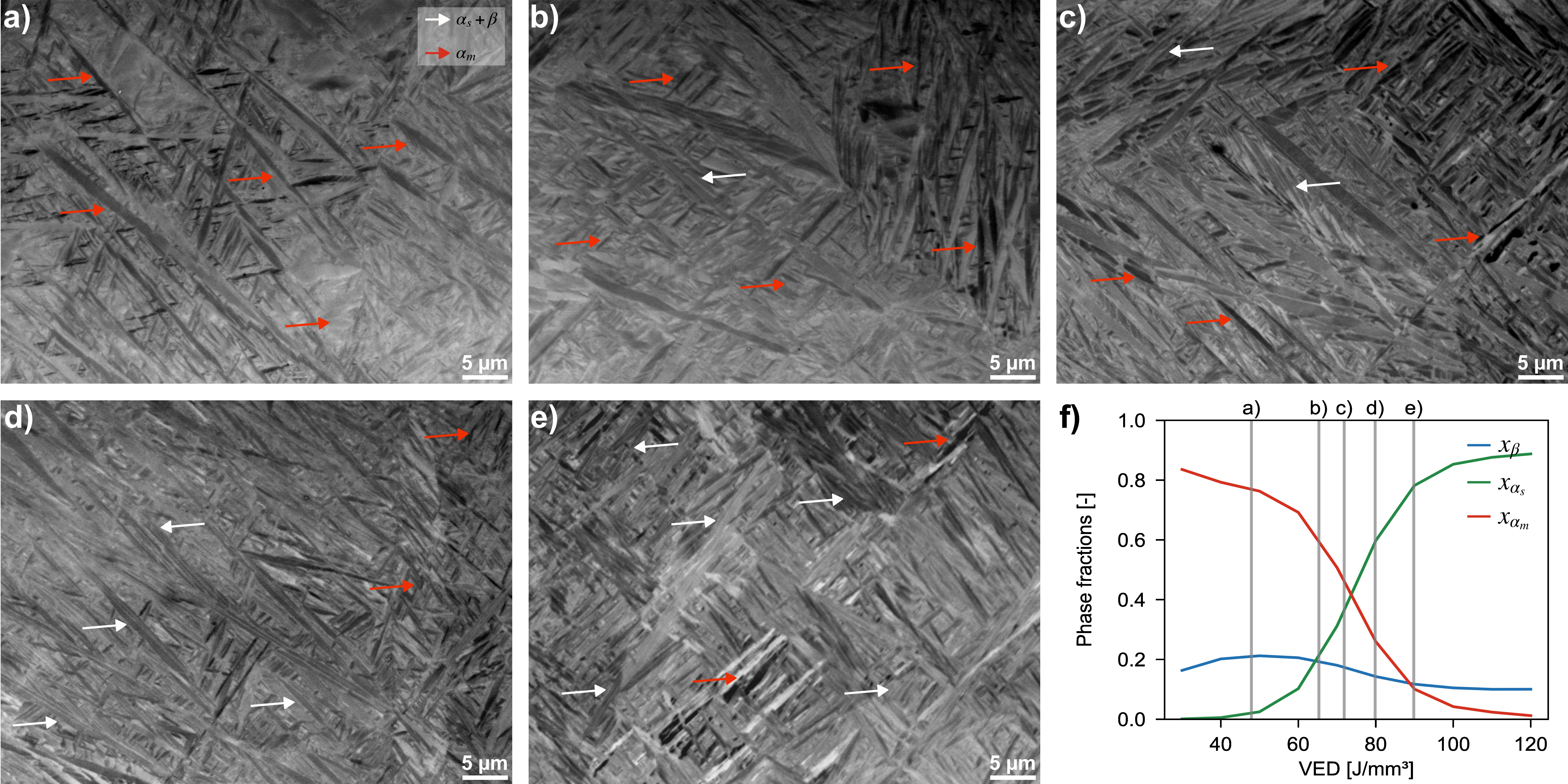}
    \caption{SEM micrographs from build job 1 of the samples with VED =
    (a) \SI{47.90}{\joule\per\milli\meter\cubed}, (b) \SI{65.32}{\joule\per\milli\meter\cubed}, (c) \SI{71.85}{\joule\per\milli\meter\cubed}, (d) \SI{79.84}{\joule\per\milli\meter\cubed}, (e) \SI{89.81}{\joule\per\milli\meter\cubed}. (f) Corresponding predicted phase fractions.}
    \label{fig:microstructures_15s}
\end{figure}

The conducted experimental study allows for the comparison of the predicted final phase compositions with experimental micrographs. To this end, \cref{fig:microstructures_15s} shows representative micrographs from samples of build job 1 manufactured at VED values of \SI{48}{\joule\per\milli\meter\cubed}, \SI{65}{\joule\per\milli\meter\cubed}, \SI{80}{\joule\per\milli\meter\cubed}, and \SI{90}{\joule\per\milli\meter\cubed}. 

In detail, at the lowest VED of \SI{48}{\joule\per\milli\meter\cubed} (\cref{fig:microstructures_15s}a), the microstructure consists almost entirely of fine, acicular $\alpha_m$ martensite, characteristic of rapid solidification and limited thermal exposure during layer consolidation. 
Increasing the energy density to \SI{65}{\joule\per\milli\meter\cubed} (\cref{fig:microstructures_15s}b) results in a partial decomposition of $\alpha_m$ martensite into a mixture of $\alpha_m$ and fine $\alpha_s+\beta$ laths. 
At \SI{71}{\joule\per\milli\meter\cubed} (\cref{fig:microstructures_15s}c) and \SI{80}{\joule\per\milli\meter\cubed} (\cref{fig:microstructures_15s}d), the transformation proceeds further yielding a microstructure composed of well-defined lamellar $\alpha_s+\beta$ colonies, alongside some remnant martensitic regions.
This suggests that the energy input is sufficient to induce partial in-situ tempering during the LPBF process.
Finally, at the highest VED of \SI{90}{\joule\per\milli\meter\cubed} (\cref{fig:microstructures_15s}e), the microstructure evolves into a fully lamellar $\alpha_s+\beta$ structure, with clearly developed $\alpha_s$ laths and continuous $\beta$ films at the colony boundaries. 
This progressive transition from martensitic to equilibrium lamellar morphology confirms the possibility of lamellar $\alpha_s+\beta$ in the as-built state. Further, it underscores the strong VED dependence of phase constitution and microstructural refinement in LPBF of Ti-6Al-4V. 

The experimentally observed microstructural transition from martensitic to lamellar $\alpha_s+\beta$ with increasing VED is mirrored in the predicted phase fraction plot obtained from thermal-history-based microstructure simulations (see \cref{fig:microstructures_15s}f). 
At the lowest energy input of \SI{48}{\joule\per\milli\meter\cubed}, the predicted phase composition is dominated by $\alpha_m$-martensite (0.77), with only a minor fraction of $\alpha_s$ (0.02) and a comparatively low fraction of $\beta$ ($\approx$ 0.21). This is consistent with the fine, fully martensitic microstructure observed experimentally. As the VED increases to an intermediate VED of \SI{65}{\joule\per\milli\meter\cubed}, the $\alpha_m$ fraction decreases to 0.59, while the $\alpha_s$ and $\beta$ fractions increase to 0.21 and 0.19, respectively, reflecting partial tempering and the initial appearance of $\alpha_s+\beta$ colonies in the micrographs.
Finally, at the highest energy density of \SI{90}{\joule\per\milli\meter\cubed}, the prediction shows a nearly $\alpha_s+\beta$ equilibrium structure, with 0.77 $\alpha_s$, 0.13 $\beta$, and only 0.10 $\alpha_m$, corresponding to the fully lamellar microstructure observed experimentally.
Overall, the simulated phase fraction trends quantitatively support the microstructural observations.

Compared to Dhiman et al.'s \cite{dhiman2024Microstructure} study, the observed general trends are the same. However, in their case, the transition towards $\alpha_s+\beta$ microstructures starts at a lower VED of \SI{55}{\joule\per\milli\meter\cubed}. In both \cite{dhiman2024Microstructure} and in the present study, a $\Delta t$ of \SI{90}{\micro\metre}, a $T_b$ of \SI{200}{\celsius} and similar ILTs of \SI{12}{\second} and \SI{15}{\second} are used. Given the similar parameters, this shift can be attributed to the support structures used by Dhiman et al., which are known to yield more heat accumulation, thus promoting $\alpha_s+\beta$ \cite{xu2017Situ}. Besides, the microstructures observed in the present study feature thin laths of thicknesses less than \SI{1}{\micro\metre}, which does not correspond to the significant microstructure coarsening reported by Medvedev et al. \cite{medvedev2025Interlayer} at ILTs below \SI{40}{\second}, which they present as a general rule. Their tested VED range is similar (\SIrange{43}{149}{\joule\per\milli\meter\cubed} to this study. However, they achieve these VEDs by comparatively slow scan speeds between \SIrange{48}{231}{\milli\meter\per\second} and (partly) thicker layer thicknesses between \SIrange{60}{300}{\micro\meter}. This strongly suggests that the used slow scan speeds lead to microstructure coarsening due to increased heat accumulation.

Next to errors stemming from the microstructure model, errors in the predicted phase evolutions can also be introduced by the thermal model via error propagation. 
Small deviations in the temperature histories can potentially shift the predicted balance between competing phase evolution pathways, such as between martensitic and non-martensitic transformations. Nevertheless, overall trends - which are the focus of this study - are preserved, albeit with possible slight shifts. The study by Klusemann et al. \cite{klusemann2018Stability} supports this notion. They investigated the robustness of the predicted phase evolutions for slightly varying ($\pm$\SI{5}{\percent}) temperature histories. These temperature histories also varied in whether the temperature exceeded $T_{\beta,trans}$ at a certain point in time. Despite these variations which did cause temporary differences in phase composition, the final phase fractions did not show significant differences. Similarly to such temperature differences in simulated temperature histories, experimental temperature measurements introduce errors depending on the techniques and resolution in time and space.
When zooming in from general trends to specific parameter combinations and part geometries, it is self-evident to switch to a more complex thermal model and include sensitivity analysis, especially near transformation boundaries.

\subsection{Trends in microstructure predictions \label{sec:trends_microstructure}}
Based on the conducted DOE, the effect of the four included process parameters VED, $T_b$, $\Delta t$ and ILT on the final phase composition of the considered bulk layer can be investigated.  \cref{fig:forward_study_overview}  in \ref{sec:appendix_A} gives an overview of this four-dimensional result space. 
In the following, we first establish the link between the final phase composition and the underlying phase evolution (\cref{sec:phase_evolution}). We then analyze the influence of individual process parameters on the achievable phase compositions using phase composition maps (\cref{sec:composition_maps}) and finally evaluate the relative importance of the process parameters for determining the microstructures using selected process parameter scenarios (\cref{sec:relative_importance}).

\subsubsection{Linking final phase composition to phase evolution \label{sec:phase_evolution}}
\begin{figure}[]
    \centering
        \includegraphics[width=\textwidth]{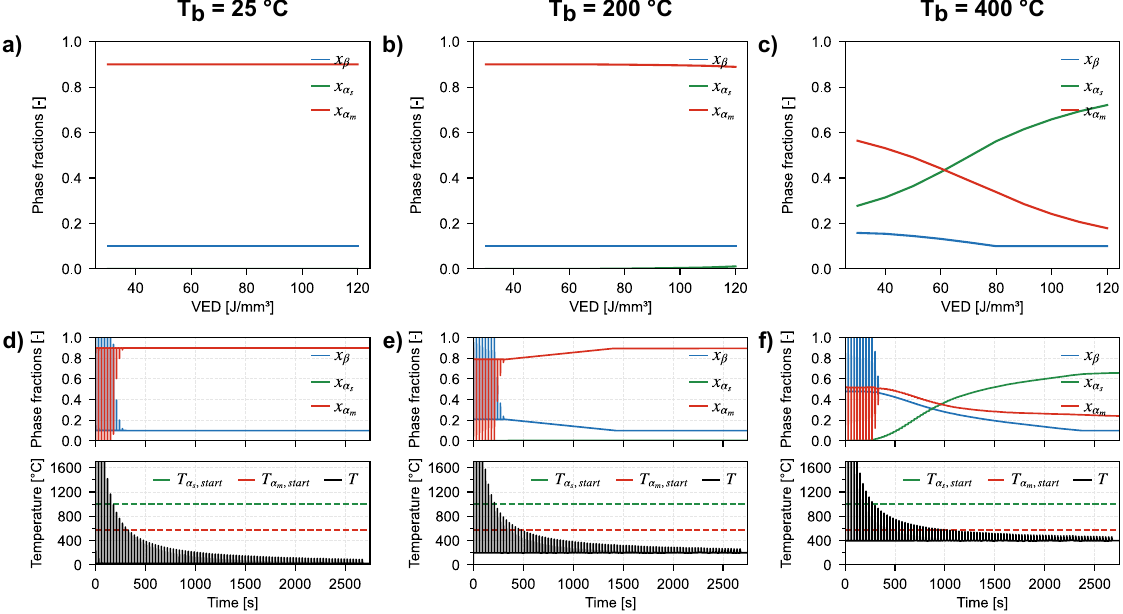}
    \caption{(Top row: Final phase composition as a function of VED at $\Delta t$ = \SI{30}{\micro\metre}, ILT = \SI{30}{s}  and $T_b$ = (a) \SI{25}{\celsius}, (b) \SI{200}{\celsius}, (c) \SI{400}{\celsius}. Bottom row: Corresponding phase evolution over time at $\Delta t$ = \SI{30}{\micro\metre}, ILT = \SI{30}{s}, a representative VED of \SI{100}{J/m^3} and a $T_b$ of (d) \SI{25}{\celsius}, (e) \SI{200}{\celsius} (f) \SI{400}{\celsius}.}
    \label{fig:30micrometer_forward_study}
\end{figure}

The analysis will first focus on a typical layer thickness for LPBF of $\Delta t$ = \SI{30}{\micro\metre}. \cref{fig:30micrometer_forward_study} shows the predicted phase fractions as a function of VED for the representative ILT = \SI{30}{s} at $T_b$ = \SI{25}{\celsius} (\cref{fig:30micrometer_forward_study}a), $T_b$ = \SI{200}{\celsius} (\cref{fig:30micrometer_forward_study}b) and $T_b$ = \SI{400}{\celsius} (\cref{fig:30micrometer_forward_study}c). For the two lower $T_b$ values, the microstructure is predicted to be fully martensitic, independently of the VED. Only at $T_b$ = \SI{400}{\celsius}, $\alpha_s$ becomes more prominent with \SI{30}{\percent} already at \SI{30}{\joule\per\milli\meter\cubed} and a steady increase to \SI{78}{\percent} at \SI{120}{\joule\per\milli\meter\cubed}.

To understand these trends, the underlying phase evolutions taking place throughout the build job are presented for the representative VED of \SI{100}{\joule\per\milli\meter\cubed} for each of the three $T_b$s in \cref{fig:30micrometer_forward_study}d, \cref{fig:30micrometer_forward_study}e and  \cref{fig:30micrometer_forward_study}f.
The corresponding temperature evolutions for a VED of \SI{100}{\joule\per\cubic\milli\meter} are given in the bottom subplots, that also indicates the temperatures relevant for phase transformations: The dashed green line marks the temperature when $\beta$ starts to transform into $\alpha_S$ and the dashed red line marks the temperature when $\beta$ starts to transform into $\alpha_m$. In addition to the temperature range below the green line, $\alpha_s$ formation is possible via $\alpha_m$ dissolution between $\approx$ \SI{450}{\celsius} and \SI{1000}{\celsius}, as indicated by $k_{\alpha_s}$ in \cref{fig:nitzler_model}c.
The three bottom subplots of \cref{fig:30micrometer_forward_study}d-f show how the temperatures shift to higher values with increasing $T_b$. From these, it becomes clear that for all $T_b$ values, the total time in the $\alpha_s$ formation regime ($\beta \rightarrow \alpha_s$) is very limited, between \SIrange{0.2}{2.5}{\second}. Similarly, for the two lower $T_b$ values, only \SIrange{0.7}{3.2}{\second} are spent in the temperature range that allows the dissolution of $\alpha_m$ into $\alpha_s$ ($\alpha_m \rightarrow \alpha_s$).
However, at $T_b$ = \SI{400}{\celsius}, the elevated temperatures throughout the build job do allow to slowly transform the majority of the $\alpha_m$ present into $\alpha_s$ via diffusion, with \SI{98}{\second} in the corresponding temperature range. 
In addition to the dissolution of $\alpha_m$ into $\alpha_s$, $\beta$ dissolves into $\alpha_s$ by diffusive transformation over the course of the build job. Thus, the comparatively high fractions $x_{\beta}$ of $\sim$\SI{40}{\percent} - \SI{50}{\percent} at $\sim$\SI{250}{s} reduce to \SI{10}{\percent} over time for all $T_b$.

\begin{figure}[]
    \centering
    \includegraphics[width=\textwidth]{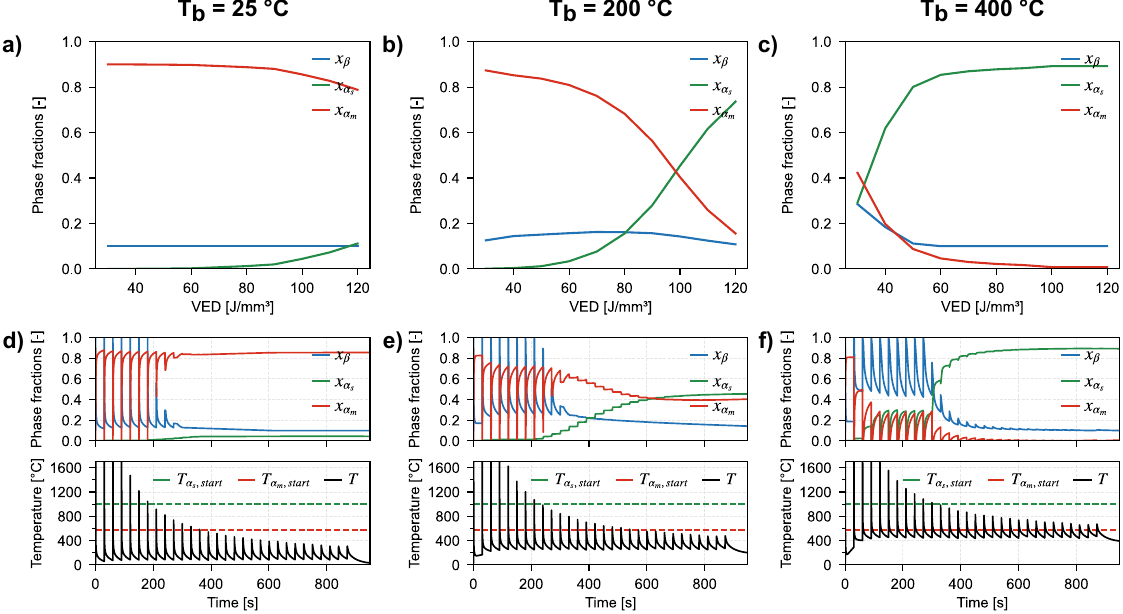}
    \caption{Top row: Final phase composition as a function of VED at $\Delta t$ = \SI{90}{\micro\metre}, ILT = \SI{30}{s}  and $T_b$ = (a) \SI{25}{\celsius}, (b) \SI{200}{\celsius}, (c) \SI{400}{\celsius}. Bottom row: Corresponding phase evolution over time at a representative VED of \SI{100}{J/m^3} and a $T_b$ of (d) \SI{25}{\celsius}, (e) \SI{200}{\celsius} and (f) \SI{400}{\celsius}.}
    \label{fig:90micrometer_forward_study}
\end{figure}

In contrast to the LPBF-typical layer thickness of $\Delta t$ = \SI{30}{\micro\metre}, a $\Delta t$ = \SI{90}{\micro\metre} offers more variety in terms of final phase compositions attainable (see  \cref{fig:90micrometer_forward_study}). The  final phase compositions as a function of VED are shown for build plate temperatures of $T_b$ = \SI{25}{\celsius} (\cref{fig:90micrometer_forward_study}a), $T_b$ = \SI{200}{\celsius} (\cref{fig:90micrometer_forward_study}b) and $T_b$ = \SI{400}{\celsius} (\cref{fig:90micrometer_forward_study}c), for an ILT = \SI{30}{s}. 
At \SI{25}{\celsius}, a VED $\leq$ \SI{80}{\joule\per\milli\meter\cubed} is predicted to yield a fully martensitic microstructure. In contrast, $x_{\alpha_s}$ continuously increases to $\geq$\SI{10}{\percent} at \SI{120}{\joule\per\milli\meter\cubed}. 
For \SI{200}{\celsius}, this shift from fully $\alpha_m$ microstructures to mixed $\alpha_m$+$\alpha_s$ microstructures takes place at a significantly lower VED of \SI{50}{\joule\per\milli\meter\cubed}. Therefore, the highest studied VED of \SI{120}{\joule\per\milli\meter\cubed} results in $x_{\alpha_s}$ $\geq$\SI{75}{\percent} compared to \SI{0}{\percent} for the same conditions at a layer thickness of \SI{30}{\micro\metre}. 
This trend is also visible at \SI{400}{\celsius}, where even the lowest studied VED of \SI{30}{\joule\per\milli\meter\cubed} produces a final phase composition of $x_{\alpha_s}$ = \SI{28}{\percent}, $x_{\alpha_m}$ = \SI{43}{\percent} and $x_{\beta}$ = \SI{28}{\percent}. Already at VED = \SI{50}{\joule\per\milli\meter\cubed}, $x_{\alpha_s}$ equals \SI{90}{\percent}.

The differences between $\Delta t$ = \SI{30}{\micro\metre} and $\Delta t$ = \SI{90}{\micro\metre} can be understood by looking at the underlying phase evolutions over time. \cref{fig:90micrometer_forward_study}d, \cref{fig:90micrometer_forward_study}e and  \cref{fig:90micrometer_forward_study}f present phase evolutions for a representative VED of \SI{100}{\joule\per\milli\meter\cubed}. First note, that the reduction in total number of layers and the associated reduction in total time from \SI{2700}{s} to \SI{950}{s} originate from the fact that the same build height is analyzed independently of the $\Delta t$ for better comparability. 
\cref{fig:90micrometer_forward_study}d for $T_b=$\SI{25}{\celsius} resembles \cref{fig:30micrometer_forward_study}d, which corresponds to the same process parameters except the layer thickness of $\Delta t$ = \SI{30}{\micro\metre}. The temperatures are above \SI{450}{\celsius} for only \SI{9}{\second} in total. Hence, $\alpha_m$ does not have time to dissolve into $\alpha_s$. 
\cref{fig:90micrometer_forward_study}e at $T_b$ = \SI{200}{\celsius} exhibits the dissolution of $\beta$ and $\alpha_m$, with $x_{\alpha_s}$ increasing over time, like in \cref{fig:30micrometer_forward_study}f, because \SI{54}{\second} are spent in the associated temperature range. 
As illustrated in \cref{fig:90micrometer_forward_study}f, $T_b$ = \SI{400}{\celsius} causes the decomposition of $\beta$ into both $\alpha_m$ and $\alpha_s$, meaning the system cycles between all three phases as long as the temperature reaches values higher than $T_{\alpha_s,start}$. Consequently, $x_{\alpha_s}$ is already approximately \SI{30}{\percent} at the end of that initial phase. Throughout the remaining build job, $\beta$ and $\alpha_m$ rapidly decompose into $\alpha_s$ resulting in a high $x_{\alpha_s}$ of \SI{90}{\percent}. The system spends \SI{38}{\second} in the green temperature regime that allows to form $\alpha_s$ from $\beta$, while this time is below \SI{6}{\second} for all other cases discussed. The time that allows $\alpha_m$ to dissolve totals \SI{811}{\second}.

\begin{figure}[]
    \centering
    \includegraphics[width=\textwidth]{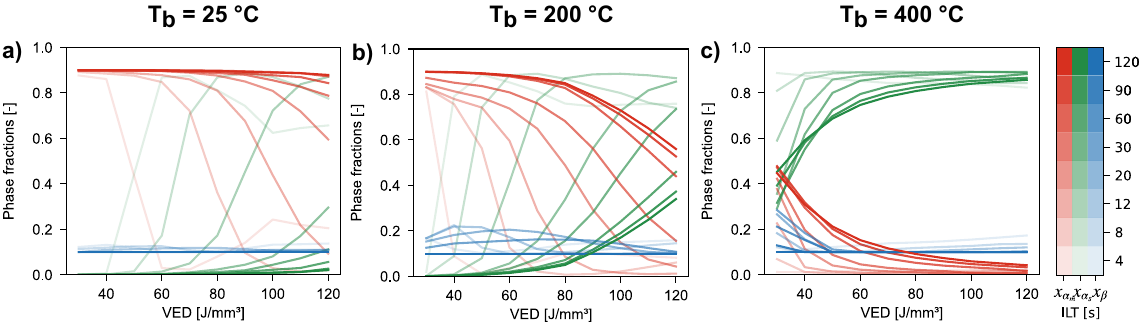}
    \caption{Final phase composition as a function of VED at $\Delta t$ = \SI{90}{\micro\metre}, and $T_b$ = (a) \SI{25}{\celsius}, (b) \SI{200}{\celsius}, and (c) \SI{400}{\celsius}, with increasing ILT denoted by more saturated colors, as indicated in the legend on the right.}
    \label{fig:90micrometer_forward_study_ILT}
\end{figure}

So far, ILT was kept constant at \SI{30}{s}. In contrast, \cref{fig:90micrometer_forward_study_ILT} visualizes the effect of varying ILTs using the example of $\Delta t$ = \SI{90}{\micro\meter} and $T_b$ = \SI{25}{\celsius} (\cref{fig:90micrometer_forward_study_ILT}a), $T_b$ = \SI{200}{\celsius} (\cref{fig:90micrometer_forward_study_ILT}b) and $T_b$ = \SI{400}{\celsius} ( \cref{fig:90micrometer_forward_study_ILT}c). 
For all $T_b$s, longer ILTs consistently shift the microstructures dominated by $\alpha_s$ to higher VED. 
Short ILT $\leq$ \SI{12}{s} are predicted to allow $x_{\alpha_m}$ < \SI{10}{\percent} even at $T_b$ = \SI{25}{\celsius}. This proves that a short ILT causes enough heat accumulation to trigger the dissolution of $\alpha_m$ and/or $\beta$. 
Medvedev et al.  \cite{medvedev2025Interlayer} found ILT effects to be independent of the part geometry, which would imply that these results are transferable to other geometries. 

Regarding the effects of the build plate temperature, Ali et al. \cite{ali2017Insitu} observed a fully $\alpha_s+\beta$ microstructure at a pre-heating temperature of \SI{570}{\celsius}. They used a $\Delta t$ of \SI{50}{\micro\meter}, a laser power of \SI{200}{\watt}, a hatch spacing of \SI{80}{\micro\meter} and exposure times of \SI{100}{\micro\second} with point distance and ILT not specified. While no direct comparison is possible due to these non-matching process parameters, the results do not contradict each other. The highest $T_b$ investigated in the present study is \SI{400}{\celsius}, for which $\alpha_m$ always significantly decomposes at a layer thickness of \SI{60}{\micro\meter}, whereas significant $\alpha_m$ decomposition occurs only for short ILTs and/or higher VED for $\Delta t=$\SI{30}{\micro\meter} (see \cref{fig:forward_study_overview} in \ref{sec:appendix_A}). 

\subsubsection{Effect of process parameters on attainable phase compositions \label{sec:composition_maps}}

\begin{figure}[]
    \centering
    \includegraphics[width=\textwidth]{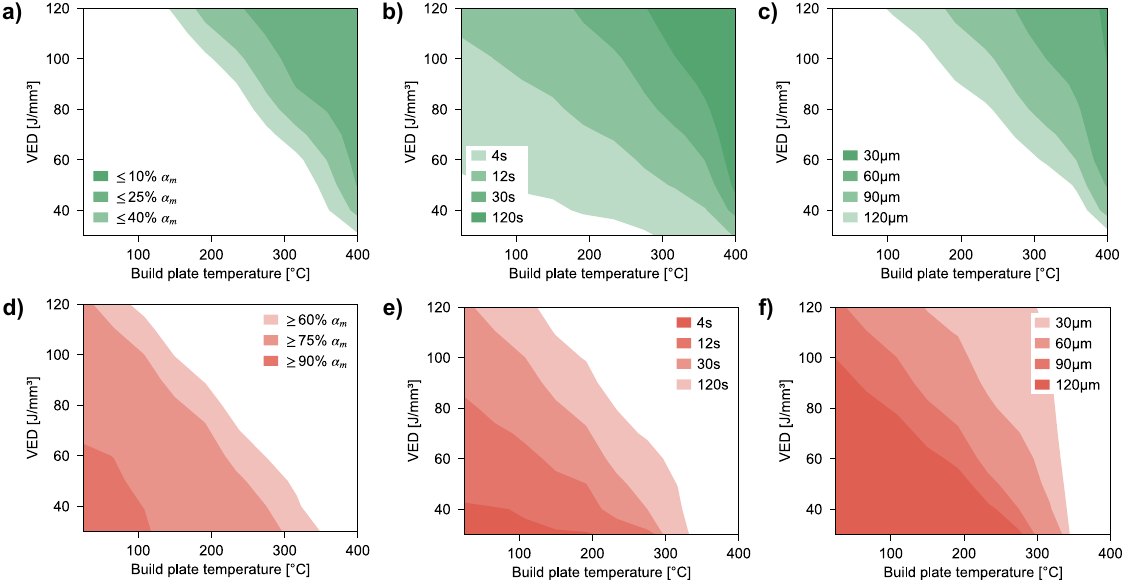}
    \caption{Maps showing regions with a phase composition of primarily $\alpha_m$  (marked in red) and primarily $\alpha_s$ + $\beta$ (marked in green) as a function of VED and $T_b$. These regions are displayed for (a,d) ILT = \SI{30}{\second}, $\Delta t$ = \SI{90}{\micro\meter}, (b,e) ILT ranging from \SI{4}{\second} to \SI{120}{\second} and $\Delta t$ = \SI{90}{\micro\meter}, and (c,f) ILT = \SI{30}{\second} and $\Delta t$ ranging from \SI{30}{\micro\meter} to \SI{120}{\micro\meter}. In (a,d) $\alpha_m$ as indicated, while $\alpha_m \leq$ \SI{25}{\percent} or $\alpha_m \geq$ \SI{75}{\percent}, respectively for the remaining plots. These contour plots are based on the original DOE, linearly interpolated onto an equidistant 10×10 grid for visualization.  
    }
    \label{fig:map_fraction}
\end{figure}

After analyzing the underlying phase transformations (\cref{sec:phase_evolution}), this section investigates how these mechanisms collectively shape the observable phase compositions across the parameter space based on phase composition maps:
The $\alpha_m$ phase content regions in \cref{fig:map_fraction} are valid for ILT = \SI{30}{\second} and $\Delta t$ = \SI{90}{\micro\meter}. The shapes of these regions underline the interchangeability of the two process parameters VED and $T_b$. Furthermore, the shapes allow to derive specific guidelines. For example, under the process conditions given above, phase compositions with $\alpha_m$ $\leq$ \SI{25}{\percent} are achieved starting from \SI{100}{\joule\per\milli\meter\cubed} at $T_b$ $\geq$ \SI{200}{\celsius}, or from \SI{80}{\joule\per\milli\meter\cubed} at $T_b$ $\geq$ \SI{300}{\celsius}. The $\alpha_m$ $\geq$ \SI{90}{\percent} region is very confined to $\approx$\SI{60}{\joule\per\milli\meter\cubed} and $\approx$\SI{100}{\celsius} as a result of the selected combination of ILT and $\Delta t$.
The other four subplots use $\alpha_m$ $\geq$ \SI{75}{\percent} and $\leq$ \SI{25}{\percent}, respectively, while varying ILT and $\Delta t$. While the general trends have already been described above, several effects can be more clearly observed here and better understood in terms of their extent:
The large green regions in \cref{fig:map_fraction}b corresponding to \SI{4}{\second} and \SI{12}{\second} clearly demonstrate how short ILTs are very effective in promoting $\alpha_s$ + $\beta$ phase compositions. Similarly, the large red regions in \cref{fig:map_fraction}f corresponding to \SI{30}{\micro\meter} and \SI{60}{\micro\meter} highlight how standard LPBF layer thicknesses consistently yield microstructures dominated by $\alpha_m$. In addition, the vertical edge of the red region associated with \SI{30}{\micro\meter} shows that setting $T_b$ $\geq$ \SI{400}{\celsius} remains a feasible method to avoid $\alpha_m$ even at thin layers of \SI{30}{\micro\meter}.
The regions indicated in \cref{fig:map_fraction} are based on the 1D FD thermal model employed throughout this study and therefore assume a simple geometry with a fixed cross-section. 

When transferring these results to more complex geometries, these regions are expected to shift. For example, heat accumulation in geometries with an increasing cross-section is assumed to shift them toward lower $T_b$ and VED due to the limited heat flux in the z-direction, whereas the opposite effect is expected for converging cross-sections. These shifts are consistent e.g. with thermal camera measurements during LPBF reported by Kavas et al. \cite{kavas2026Layer}, who observed distinct increases and decreases in surface temperature on a totem geometry depending on the local cross-section area trajectory, suggesting a similar trend within the bulk of the part. Furthermore, when the thickness of a part or feature is significantly reduced (even while maintaining the in-plane aspect ratio), the limited heat loss to the powder bed becomes more dominant and is likewise expected to shift the regions toward lower $T_b$ and VED. Support structures follow the same logic, as they limit the heat loss to the build plate.

\subsubsection{Relative importance of process parameters \label{sec:relative_importance}}

\begin{figure}[]
    \centering
    \includegraphics[width=\textwidth]{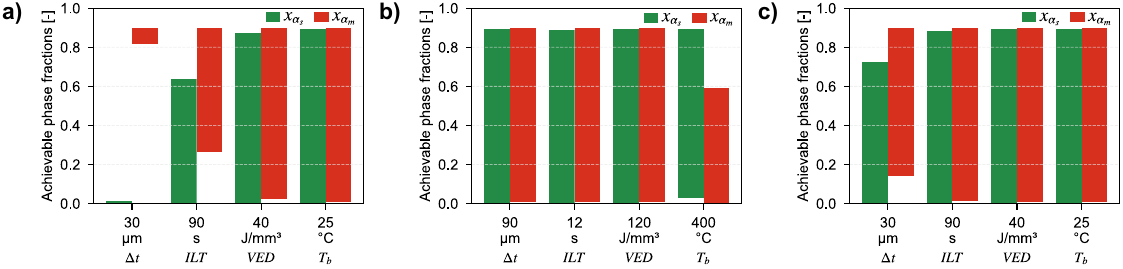}
    \caption{Achievable phase fraction ranges for different process parameters restrictions. The restricted parameter is indicated on the x-axis for each set of bars. Compilation of process parameter restrictions yielding 
    (a) a low-heat-accumulation environment, (b) a high-heat-accumulation environment, (c) a low-heat-accumulation environment. 
    For (a) and (b), $T_b$ is fixed to \SI{200}{\celsius} for all bars but the fourth bar to increase the informational value, while (c) lifts this restriction. 
    }
    \label{fig:achievable_phase_fractions_combined}
\end{figure}

To assess the relative influence of individual process parameters on the achievable microstructures, \cref{fig:achievable_phase_fractions_combined} illustrates the maximum achievable range of $\alpha_s$ and $\alpha_m$ fractions according to the predictions when varying all but one of the key parameter VED, $T_b$, $\Delta t$, and ILT: In \cref{fig:achievable_phase_fractions_combined}a, the one process parameter that is not varied is set to a value that rather promotes a low-heat-accumulation-accumulation environment, e.g. VED $ \SI{40}{\joule\per\milli\meter\cubed}$. In contrast, in \cref{fig:achievable_phase_fractions_combined}b, the not varied process parameter is set to a value rather promoting a high-heat-accumulation environment, e.g. VED $ \SI{120}{\joule\per\milli\meter\cubed}$. To not obscure the effect of VED, $\Delta t$ and ILT by the persistent effect of $T_b$ = \SI{400}{\celsius}, the $T_b$ is fixed at \SI{200}{\celsius} in \cref{fig:achievable_phase_fractions_combined}a and \cref{fig:achievable_phase_fractions_combined}b. \cref{fig:achievable_phase_fractions_combined}c equals \cref{fig:achievable_phase_fractions_combined}a but lifts that $T_b$ restriction. For each parameter, the dataset was screened to identify the combinations that yield the lowest and highest $\alpha_s$ and $\alpha_m$ fractions, thereby revealing the sensitivity of phase evolution to process conditions.

In the low-heat-accumulation environment of \cref{fig:achievable_phase_fractions_combined}a, at a layer thickness of \SI{30}{\micro\metre}, the $\alpha_s$ fraction varies only modestly between 0.00 and 0.01, whereas $\alpha_m$ remains dominant, ranging between 0.82 and 0.90, suggesting that thin layers promote rapid cooling and preserve the martensitic microstructure. For an ILT of \SI{90}{s}, the $\alpha_s$ fraction increases to 0.00–0.64, and $\alpha_m$ decreases to 0.26–0.90, indicating that prolonged cooling intervals promote martensite.  VED = \SI{40}{\joule\per\milli\meter\cubed} or $T_b$ = 25 °C lead to $\alpha_s$ fractions from 0.00 to 0.87/0.89 and $\alpha_m$ from 0.02/0.01 to 0.90 covering almost the entire ranges. 
The analysis highlights that in this low-heat-accumulation environment $\Delta t$ and ILT have the strongest effect on phase constitution and the values set for them limit the range of microstructures that can form, whereas VED and $T_b$ play secondary, moderating roles by influencing the cooling rate and heat accumulation.

\cref{fig:achievable_phase_fractions_combined}b illustrates the predicted ranges of $\alpha_s$ and $\alpha_m$ fractions for a second set of process conditions, again with $T_b$ fixed at \SI{200}{\celsius}. In contrast to the previous, lower-heat parameter set, these conditions represent higher thermal inputs - specifically, increased layer thickness (\SI{90}{\micro\meter}), short ILT (\SI{12}{\second}), and high volumetric energy density (\SI{120}{\joule\per\milli\meter\cubed}). Across these three parameters, the predicted $\alpha_s$ and $\alpha_m$ fractions exhibit very broad ranges - $\alpha_s$ between 0.00–0.89/0.90, $\alpha_m$ between 0.00/0.01–0.90 - indicating that such high-energy or thermally accumulative conditions expand the accessible microstructural design space. This implies that thicker layers, shorter ILTs, and higher VEDs enhance the tunability of the phase constitution by providing more flexible thermal profiles during solidification and cooling.
In contrast, variation in $T_b$ shows a more directional effect. At \SI{400}{\celsius}, the predicted $\alpha_s$ fraction ranges from 0.03–0.89, while $\alpha_m$ decreases substantially to 0.00 up to only 0.59, confirming that elevated substrate preheating suppresses martensitic transformation and promotes equilibrium $\alpha_s+\beta$ formation.
These results highlight that while higher energy and shorter cycle parameters (large layer thickness, short ILT, high VED) broaden the achievable phase composition space, increased $T_b$ narrows it down favoring lamellar $\alpha_s+\beta$ microstructures.

\cref{fig:achievable_phase_fractions_combined}c illustrates the predicted $\alpha_s$ and $\alpha_m$ phase fraction ranges when the restriction to a fixed $T_b$ of \SI{200}{\celsius} from \cref{fig:achievable_phase_fractions_combined}a is lifted. The resulting parameter space shows significantly broader phase variability compared to \cref{fig:achievable_phase_fractions_combined}a.
For a layer thickness of \SI{30}{\micro\meter}, $\alpha_s$ spans 0.00–0.72 and $\alpha_m$ 0.14–0.90, indicating that variable substrate preheating substantially widens the accessible range of phase composition. The combined results for ILT and VED further show $\alpha_s$: 0.00–0.88/0.89 and $\alpha_m$: 0.01–0.90, mirroring the trends observed previously, but now extending over a larger design space.
These findings highlight that $T_b$ serves as an effective compensating parameter, fully counteracting the restrictive behavior seen for ILT and VED in \cref{fig:achievable_phase_fractions_combined}a. This underscores the strong moderating role of substrate preheating in tailoring microstructural flexibility in LPBF of Ti-6Al-4V.

Based on the gained understanding of the interplay of the investigated four key process parameters, the following design guidelines  can be derived: If short ILTs are feasible (due to e.g., limited number of parts per build plate and fast recoating times), recommended process settings to reduce $\alpha_m$ are $\Delta t = \SI{60}{\micro\meter}-\SI{90}{\micro\meter}$, medium VED, and $T_b = \SI{200}{\celsius}$. When short ILTs are not possible, a martensitic microstructure can still be avoided by increasing either VED or $T_b$ to compensate for reduced heat accumulation. If high spatial resolution is not required, thicker $\Delta t$ $\approx \SI{90}{\micro\meter}$ should be preferred for maximizing microstructural tunability and higher flexibility in selecting other process parameters. The VED serves as a locally adjustable parameter that fine-tunes microstructure once global settings are fixed. Higher VED raises peak temperatures and enables greater fractions of $\alpha_s$, offering a convenient lever to locally tailor microstructures within the same build job or to compensate for constraints imposed by ILT and $\Delta t$. Finally, $T_b$ acts as a global compensation parameter. Elevated $T_b$ values can offset limitations from other parameters, allowing access to nearly the full range of $\alpha_s$ fractions - even for long ILTs (e.g., $\SI{90}{\second}$), low VED (\SI{40}{\joule\per\milli\meter\cubed}), and thin $\Delta t$ (\SI{30}{\micro\meter}).

\subsection{Applying the generated process map to practical applications \label{sec:practical_applications}}

\subsubsection{Compensating for increasing ILT}

A common scenario in practice is an increase in number of parts on the build plate due to changing requirements e.g. when switching from a preliminary study to production. Typically, the same process parameters are used despite the inevitably increasing ILT. Hence, the as-built microstructure changes in an uncontrolled way.

\begin{figure}[t!]
    \centering
    \includegraphics[width=\linewidth]{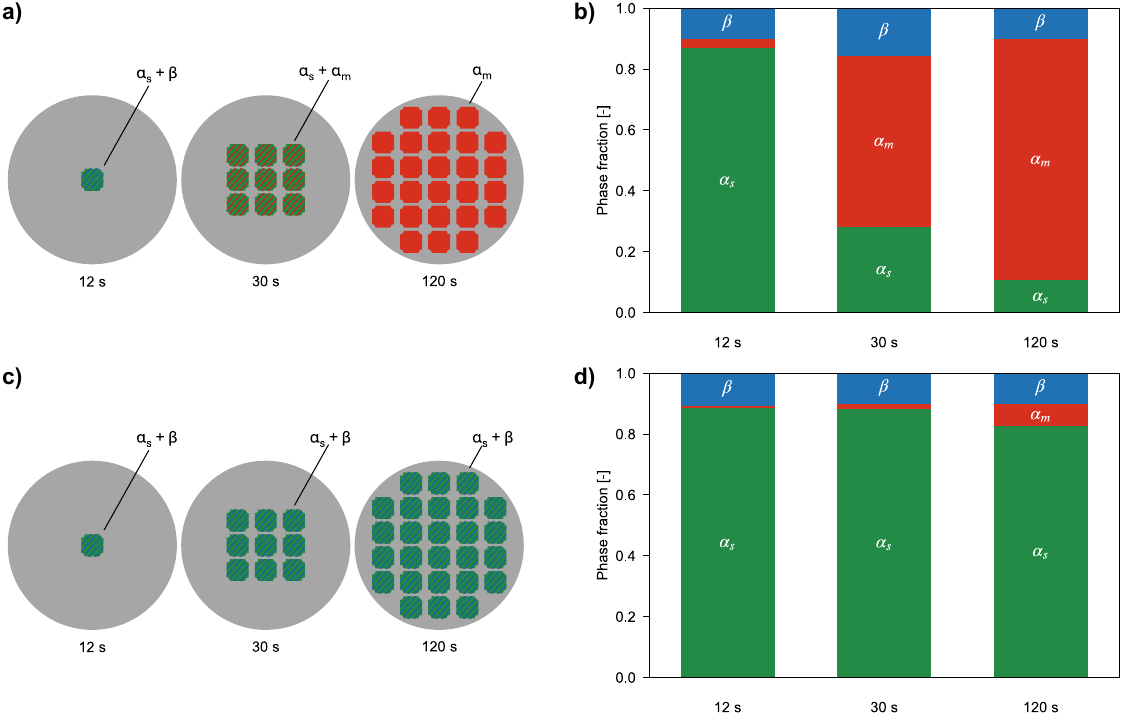}
    \caption{Schematic drawing of microstructures (a) changing and (c) remaining consistent with increasing ILT. Final phase composition depending on ILT for VED = \SI{90}{\joule\per\milli\meter\cubed}, $\Delta t$ = \SI{90}{\micro\metre} (b) \SI{200}{\celsius} and (d) \SI{400}{\celsius}.}
    \label{fig:case_study_ILT}
\end{figure}

As an example, we assume that the process parameters were selected to achieve an $\alpha_s+\beta$ microstructure in the as-built state for a build job consisting of few parts and hence a short ILT of \SI{12}{\second}.
For this scenario, one possible set of process parameters is $T_b = \SI{200}{\celsius}$, $\Delta t = \SI{90}{\micro\meter}$, and ILT $= \SI{12}{\second}$. As displayed in \cref{fig:case_study_ILT}a, those process parameters yield \SI{90}{\percent} $\alpha_s$ and \SI{10}{\percent} $\beta$.
As the ILT rises to \SI{30}{\second}, and eventually \SI{120}{\second}, the as-built microstructure gradually shifts from a lamellar $\alpha_s+\beta$ morphology to a mixed $\alpha_s$ and $\alpha_m$ microstructure, and finally to a fully martensitic $\alpha_m$ microstructure (see  \cref{fig:case_study_ILT}b). This behavior reflects the enhanced cooling between layers that occurs when longer ILTs interrupt the heat accumulation typical for smaller build jobs.

However, the conducted parameter sweep demonstrates that these undesired effects can be compensated through adjustments in other process parameters, enabling restoration of the target lamellar $\alpha_s+\beta$ morphology as displayed in \cref{fig:case_study_ILT}c. Three compensation strategies were evaluated.

\begin{figure}[]
    \centering
    \includegraphics[width=\textwidth]{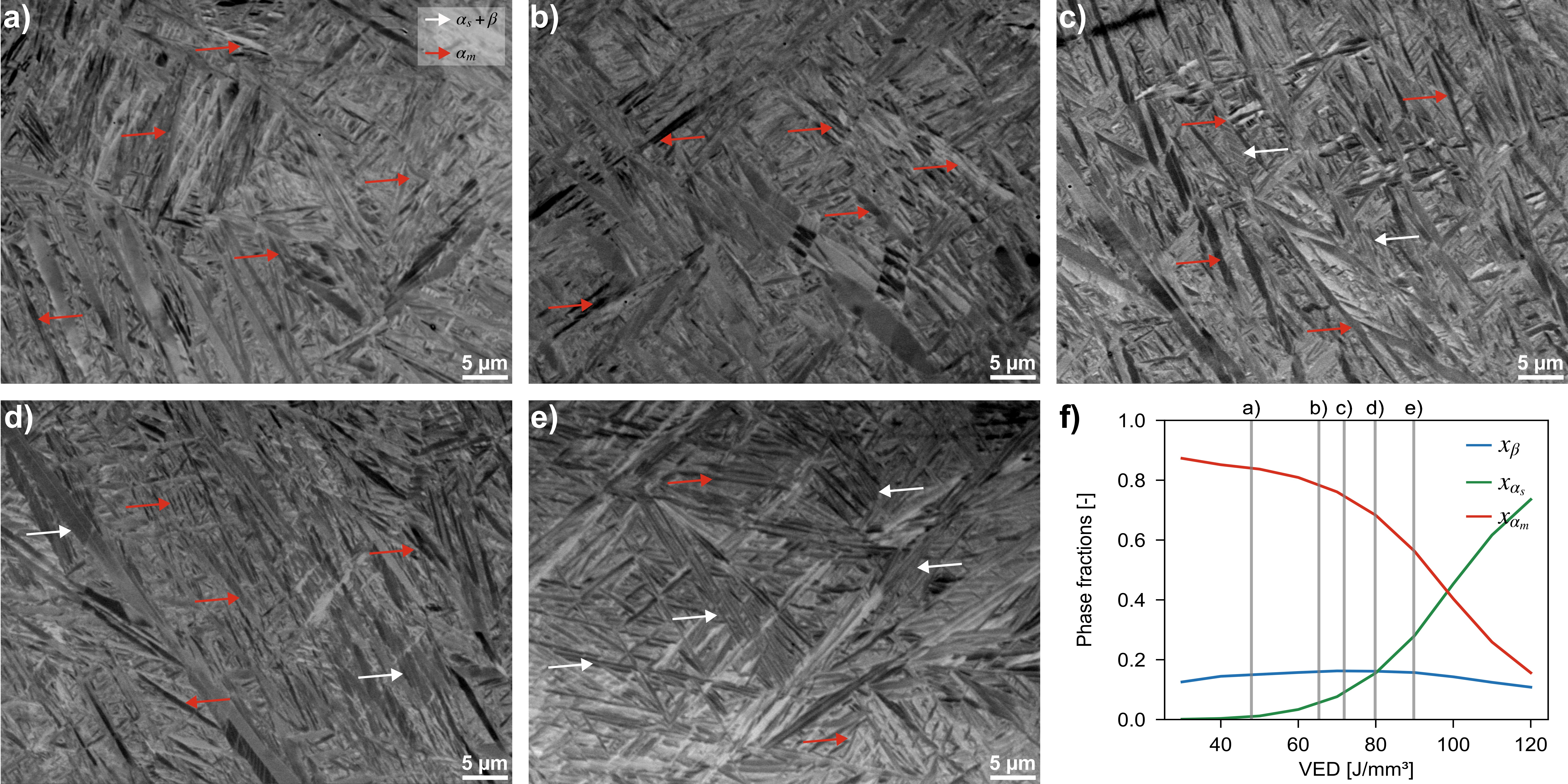} 
    \caption{SEM micrographs from build job 2 of the samples with VED =
    (a) \SI{47.90}{\joule\per\milli\meter\cubed}, (b) \SI{65.32}{\joule\per\milli\meter\cubed}, (c) \SI{71.85}{\joule\per\milli\meter\cubed}, (d) \SI{79.84}{\joule\per\milli\meter\cubed}, (e) \SI{89.81}{\joule\per\milli\meter\cubed}. (f) Corresponding predicted phase compositions.}
    \label{fig:microstructures_30s}
\end{figure}

The first strategy compensates for the increased ILT solely by raising the VED, without changing other process parameters. This approach is effective for moderate ILTs - specifically up to around \SI{30}{\second} - but becomes infeasible for substantially longer times due to process instability. To experimentally validate this approach, two build jobs were conducted at ILTs of \SI{15}{\second} and \SI{30}{\second}. The accompanying micrographs for the \SI{12}{\second} case were discussed in \cref{sec:accuracy_microstructure}, while those for the \SI{30}{\second} condition are presented in \cref{fig:microstructures_30s}.
The micrographs reveal a similar microstructural trend to that observed at \SI{15}{\second}: At the lowest VED of \SI{48}{\joule\per\milli\meter\cubed} (\cref{fig:microstructures_30s}a), the microstructure is dominated by fine, acicular $\alpha_m$ martensite, reflecting the high cooling rates and limited thermal exposure during consolidation.
Increasing the VED to \SI{65}{\joule\per\milli\meter\cubed} (\cref{fig:microstructures_30s}b) initiates the first signs of martensite decomposition, leading to a microstructure that still consists predominantly of $\alpha_m$ but with emerging tempered regions.
At \SI{71}{\joule\per\milli\meter\cubed} (\cref{fig:microstructures_30s}c) and \SI{80}{\joule\per\milli\meter\cubed} (\cref{fig:microstructures_30s}d), discernible lamellar features arise, marking the onset of $\alpha_s+\beta$ formation while portions of $\alpha_m$ remain.
At the highest VED of \SI{90}{\joule\per\milli\meter\cubed} (\cref{fig:microstructures_30s}e), the microstructure transitions into a fully lamellar $\alpha_s+\beta$ structure, characterized by robust $\alpha_s$ laths and continuous $\beta$ films outlining the colony boundaries. 
The corresponding predictions (see \cref{fig:microstructures_30s}f) approximate this change from $\alpha_m$ to $\alpha_s+\beta$, whose onset is shifted toward higher VED compared to \SI{15}{s}. 
In terms of the corresponding experimental results, it should be first noted that LPBF micrographs generally show local variations in microstructure arising e.g. from dynamic melt pool phenomena or localized cooling rate fluctuations. These local differences cannot be dissolved by a 1D thermal model. As such, it is very difficult to clearly identify that the onset of the lamellar $\alpha+\beta$ microstructure is shifted toward higher VED values for \SI{30}{\second} compared to \SI{15}{\second}. However, the micrographs do not contradict the predictions of the proposed framework and remain consistent with the overall process–structure relationships derived from the simulations.

The second strategy to mitigate $\alpha_m$ microstructures involves raising $T_b$ to \SI{400}{\celsius} yielding the illustration in \cref{fig:case_study_ILT}c. For these updated process parameters of the second strategy, the final phase fractions are plotted in \cref{fig:case_study_ILT}d. For the considered ILTs of \SI{12}{\second}, \SI{30}{\second} and \SI{120}{\second}, a fully lamellar $\alpha+\beta$ structure is restored matching the schematic drawing in \cref{fig:case_study_ILT}c by reducing the cooling rate and thus promoting diffusive phase transformations.

A third strategy applies a combined adjustment of parameters: According to the generated process map, increasing VED to \SI{110}{\joule\per\milli\meter\cubed} and $T_b$ to \SI{300}{\celsius} restores the $\alpha_s+\beta$ microstructure. This strategy is limited to intermediate ILTs of e.g. \SI{60}{\second} as $T_b$ is only increased to \SI{300}{\celsius}, but maintains more flexibility as none of the parameters are set to extreme values.

According to the framework's predictions, all three strategies successfully reestablish the desired lamellar $\alpha+\beta$ morphology, yet they present different trade-offs. The higher build plate temperature approach (second strategy) offers a direct and robust solution that is known for reducing residual stresses \cite{vrancken2016Preheating}, but not all commercial LPBF machines offer build plate heating, especially above \SI{200}{\celsius} \cite{ali2017Insitu}. Solely increasing VED (first strategy) as a compensation mechanism is feasible up to moderate ILTs. The combined adjustment approach (third strategy) provides more flexibility in process parameter and is better suited for complex builds where local heat effects dominate $T_b$. Together, these findings demonstrate how multi-parameter compensation enables microstructure tailoring even under changing build conditions such as varying ILTs or build plate usage.

\subsubsection{Varying microstructure with build height}

\begin{figure}
    \centering
    \includegraphics[width=\linewidth]{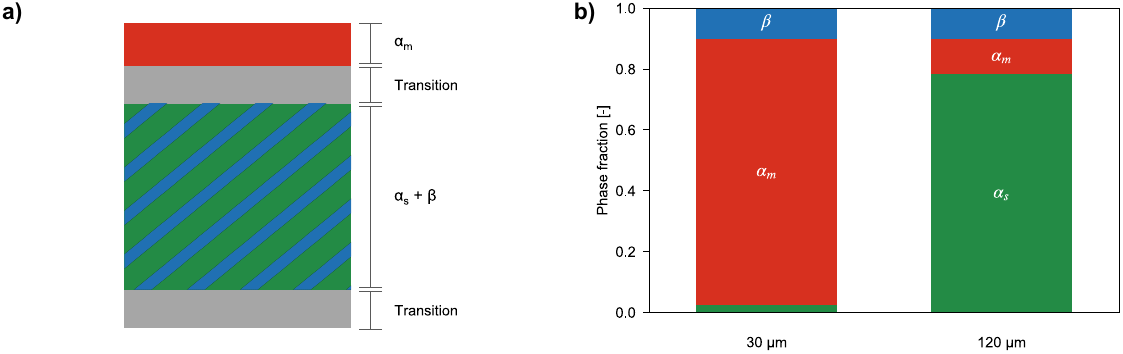}
    \caption{(a) Schematic drawing of a part with varying microstructure over the build height.
    (b) Final phase composition depending on $\Delta t$ at $T_b = \SI{300}{\celsius}$,
    VED $= \SI{90}{\joule\per\milli\meter\cubed}$, ILT $= \SI{60}{s}$.}
    \label{fig:case_study_build_height}
\end{figure}

A further step toward functional material design in LPBF is the intentional variation of microstructure along the build height, enabling the fabrication of components with wear-resistant martensitic surfaces and ductile $\alpha_s+\beta$ cores, as illustrated in \cref{fig:case_study_build_height}a. Reported hardness differences of roughly \SI{50}{HV 0.05} between martensitic and lamellar $\alpha+\beta$ microstructures \cite{barriobero-vila2017Inducing} underscore the functional advantages of such gradients.
This approach 
serves as a precursor to the heterogeneous microstructure concepts discussed in the next \cref{sec:application_VED}. 

For this application, we assume that a martensitic microstructure formed in the bulk also manifests at the top surface when identical process parameters are applied. This assumption matches the experimental results in Barriobero-Vila et al. \cite{barriobero-vila2017Inducing} and predictions by Lee et al. \cite{lee2024Extreme}, who both observed distinct martensitic top layers despite $\alpha+\beta$ in the bulk. 
With this set up, the challenge becomes finding a feasible process path to induce a controlled transition from $\alpha_s+\beta$ to $\alpha_m$ within a single build. According to the generated process map (see \cref{fig:forward_study_overview} in \ref{sec:appendix_A}), one promising route is to exploit $\Delta t$ as the main control parameter. For example, as illustrated in \cref{fig:case_study_build_height}b, the microstructure can be shifted from lamellar $\alpha_s+\beta$ to predominantly $\alpha_m$ simply by changing $\Delta t$ from \SI{90}{\micro\meter} (yielding $\alpha_s+\beta$) to \SI{30}{\micro\meter} (yielding $\alpha_m$) at a constant VED of \SI{90}{\joule\per\milli\meter\cubed}, $T_b = \SI{300}{\celsius}$, and ILT $= \SI{60}{\second}$.

While this approach offers a straightforward means of controlling microstructure with build height, it is limited by the assumption of similar thermal conditions to the bulk layer at a build height of \SI{10.85}{mm} investigated in this study that the generated process map assumes. However, varying $\Delta t$ changes these thermal conditions and layers close to the build plate experience different thermal conditions even with fixed process parameters throughout the build job, as it acts as a large heat sink influencing cooling rates. As a result, a dedicated strategy for varying process parameters dynamically throughout the build job - for example, via an inverse gradient-based design framework - is required to fully realize graded microstructures along the vertical direction. Such an approach would enable the tailored production of parts combining surface wear resistance at the top and bottom surface with core ductility, advancing the realization of functionally graded Ti-6Al-4V and related alloys produced by LPBF.

\subsubsection{Achieving parts with varying microstructures in a single build job \label{sec:application_VED}}

\begin{figure}
    \centering
    \includegraphics[width=\linewidth]{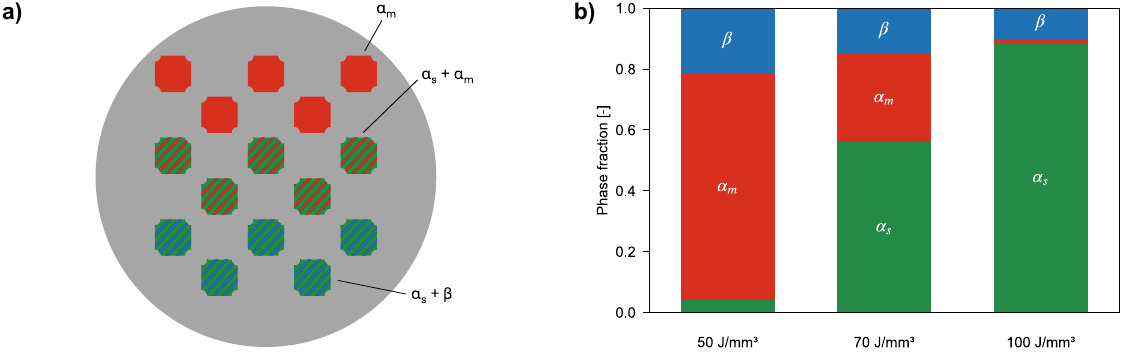}
    \caption{(a) Schematic drawing of a build job of parts with varying microstructure. (b) Final phase composition depending on VED at $T_b$ = \SI{200}{\celsius}, VED = \SI{90}{\joule\per\milli\meter\cubed}, ILT = \SI{15}{s}.}
    \label{fig:case_study_neigh_parts}
\end{figure}

A practically important and application-driven challenge is to produce adjacent parts with intentionally different as-built microstructures within the same build job, as schematically illustrated in \cref{fig:case_study_neigh_parts}a.
Such capability signifies again one important step towards parts with locally varying microstructures, like a gear wheel with a ductile $\alpha+\beta$ core and a wear-resistant martensitic ($\alpha_m$) shell, or lattices tailored e.g. for energy absorption by assigning distinct microstructures to different struts. The chosen set-up is a simplified version of such parts that is compatible with the employed 1D thermal model, as lateral heat flow within one part or lattice is not significant in this chosen set-up.

Achieving adjacent parts with distinct microstructures is nontrivial because three of the four considered key process parameters - $T_b$, $\Delta t$, and ILT - are global (job-level or at least layer-level) settings and cannot be varied independently for neighboring parts. Consequently, these parameters must be chosen so that local microstructure tuning can be accomplished by changing only VED via laser power, scan speed or hatch adjustments at the part or region level.
Consulting the generated process map (see \cref{fig:forward_study_overview} in \ref{sec:appendix_A}), one option to actualize the schematic in \cref{fig:case_study_neigh_parts}a is to set the three fixed process parameters to $T_b = \SI{200}{\celsius}$, $\Delta t = \SI{90}{\micro\meter}$, ILT $= \SI{15}{\second}$. By then selecting e.g. VEDs of \SI{50}{\joule\per\milli\meter\cubed}, \SI{70}{\joule\per\milli\meter\cubed}, and \SI{100}{\joule\per\milli\meter\cubed}  (see \cref{fig:case_study_neigh_parts}b), the process window spans from near-fully martensitic (less than \SI{10}{\percent} $\alpha_s$) to near-fully lamellar ($\approx$\SI{90}{\percent} $\alpha_s$), with intermediate VED producing mixed $\alpha_s+\alpha_m$ microstructures.

This demonstrates that, with appropriate global parameter settings, VED modulation alone is sufficient to realize distinct local microstructures in simple geometries and layouts, as could also be seen in \cref{sec:accuracy_microstructure}.
When handling localized microstructures in parts with geometric complexity, higher-fidelity (scan-resolved or 3D) simulations are required to e.g. account for lateral heat flow.

\subsubsection{Possible extension}
As the practical scenarios discussed above illustrate, a given challenge in LPBF can often be addressed through multiple processing strategies. While certain constraints may clearly favor one approach over another, in many cases several parameter combinations produce the same phase composition, leaving the choice partly arbitrary. An $\alpha$ lath thickness model would be a valuable addition to the existing framework. Such a model would introduce another level of microstructural information, enabling differentiation between fine and coarse $\alpha_s$ laths. This additional descriptor would make it possible to navigate among multiple valid parameter sets leading to the same final phase composition and to select the most suitable one based on targeted mechanical properties. For instance, finer $\alpha$ laths could be favored in regions where strength and fatigue resistance are critical. Integrating such a model would thus contribute to bridging the gap between process–structure predictions and process–structure–property optimization.

\section{Conclusion}
In this study, we developed a computational framework that couples a fast 1D thermal model with a microstructure evolution model to efficiently predict the phase fractions of $\alpha_s$, $\alpha_m$, and $\beta$ across a wide LPBF process window. By screening 2,000 parameter combinations and validating key trends against experiments, we demonstrated how this approach can map attainable microstructures, extract process–structure relationships, and support practical LPBF scenarios including microstructure preservation and spatial tailoring. Specifically, the following conclusions could be drawn:

\begin{enumerate}
    \item For the constant-cross-section geometries considered here, the coupled 1D thermal–microstructure framework was found to be a pragmatic enabler of broad process-space exploration. The 1D thermal model reproduced the relevant thermal histories with good accuracy for the thin-wall and cuboid cases. Long-term peak temperatures showed relative absolute errors between <\SI{0.1}{\percent} and \SI{6}{\percent} and cooling periods with \SIrange{0.7}{2.0}{\percent}. Importantly, experimental microstructure observations further support the predictions. At the same time, the computational cost is reduced by about $10^{3}$ relative to the 3D layer-wise FEM model and by about $3\times10^{4}$ relative to the scan-resolved 3D model. 

    \item Across the 2,000-condition DOE, the transition from $\alpha_m$-dominated to $\alpha_s + \beta$-dominated microstructures is governed by the combined thermal exposure produced by all four studied parameters, rather than by VED alone:
    \begin{itemize}
        \item Layer thickness and interlayer time are the strongest controls of final phase composition because they most strongly change cumulative thermal exposure. Thin layers such as \SI{30}{\micro\metre} favor predominantly martensitic microstructures across much of the process window, while thicker layers, especially \SI{90}{\micro\metre}, strongly expand the region where $\alpha_s + \beta$ could form. Short interlayer times, especially $\leq$\SI{12}{\sec}, promote heat accumulation and enabled strong $\alpha_m$ decomposition even at relatively low build plate temperature.
        \item Build plate temperature acts as a global compensation parameter: increasing it shifts the process window toward $\alpha_s + \beta$, and \SI{400}{\celsius} suppresses martensite even in cases that were otherwise martensitic.
        \item VED mainly serves as a local tuning variable that shifts phase boundaries within a given global thermal setting. 
    \end{itemize}

    \item The developed framework was shown to be practically useful for identifying compensation strategies in three cases. First, when interlayer time increases, the loss of heat accumulation and the resulting shift toward martensite can be compensated by increasing VED, by increasing build plate temperature, or by combining both. Second, variation of microstructure with build height appears feasible in principle, for example by changing layer thickness, but would require dynamic parameter adaptation because thermal conditions vary during the build. Third, distinct neighboring-part microstructures within one build job can be achieved when global parameters are chosen appropriately and VED is then varied locally.
\end{enumerate}  

Thus, the framework is suitable for rapid parameter selection and compensation planning, but its present predictive output is limited to phase fractions; it does not yet distinguish among parameter sets that yield similar phase fractions but different lath morphology or mechanical properties. Hence, a natural extension of this framework is the integration of a lath-thickness model.

\appendix

\section{Final phase composition as a function of key processing parameters \label{sec:appendix_A}}

\renewcommand{\thefigure}{A\arabic{figure}}
\setcounter{figure}{0}

\begin{figure}[H]
    \centering
    \includegraphics[width=1\linewidth]{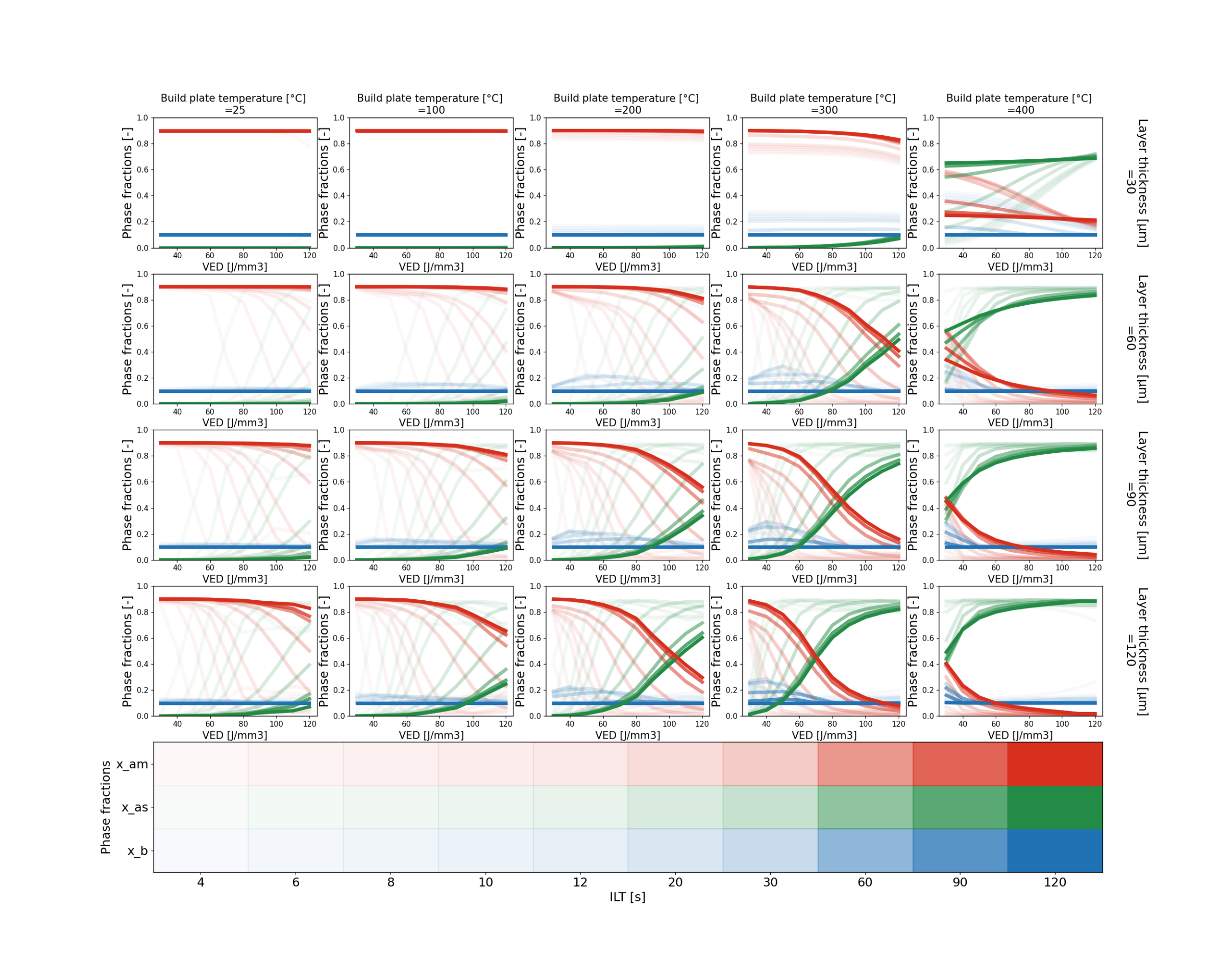}
    \caption{Final phase composition as a function of VED. $T_b$ is indicated per column, $\Delta t$  is indicated per row, ILT is decoded by color saturation according to the legend given at the bottom of the figure.}
    \label{fig:forward_study_overview}
\end{figure}

\renewcommand{\thefigure}{\arabic{figure}}

\section*{Author contributions}
\textbf{CVDL}: Conceptualization, Data Curation, Investigation, Methodology, Formal Analysis, Software, Validation, Visualization, Writing -- Original Draft.
\textbf{IS}: Methodology, Conceptualization, Formal Analysis.
\textbf{LD}: Supervision, Conceptualization, Writing -- Review \& Editing.
\textbf{MA}: Supervision, Conceptualization, Writing -- Review \& Editing.
\textbf{MB}: Funding Acquisition, Supervision, Conceptualization, Writing -- Review \& Editing.

\section*{Data Availability}
The code of the microstructure design framework is available under \url{https://github.com/C-vdL/microstructure_engineering_lpbf_Ti6Al4V}. The numerical simulation data required to reproduce these findings can be shared upon reasonable request.

\section*{Conflict of interest}
The authors declare no conflict of interest

\section*{Declaration of generative AI and AI-assisted technologies in the writing process}
During the preparation of this work, the authors used ChatGPT in order to identify typos and improve language/readability. After using this tool, the authors reviewed and edited the content as needed and take full responsibility for the content of the publication.



\bibliography{Bibliothek}

@article{ahmed1998Phase,
  title = {Phase Transformations during Cooling in {\emph{{$A$}}}+{\emph{{$\beta$}}} Titanium Alloys},
  author = {Ahmed, T. and Rack, H. J.},
  year = {1998},
  month = mar,
  journal = {Materials Science and Engineering: A},
  volume = {243},
  number = {1},
  pages = {206--211},
  issn = {0921-5093},
  doi = {10.1016/S0921-5093(97)00802-2},
  urldate = {2025-01-15}
}

@article{ali2017Insitu,
  title = {In-Situ Residual Stress Reduction, Martensitic Decomposition and Mechanical Properties Enhancement through High Temperature Powder Bed Pre-Heating of {{Selective Laser Melted Ti6Al4V}}},
  author = {Ali, Haider and Ma, Le and Ghadbeigi, Hassan and Mumtaz, Kamran},
  year = {2017},
  month = may,
  journal = {Materials Science and Engineering: A},
  volume = {695},
  pages = {211--220},
  issn = {0921-5093},
  doi = {10.1016/j.msea.2017.04.033},
  urldate = {2025-11-24}
}

@article{avrami1941kinetics,
  title = {Kinetics of Phase Change. {{III}}, Granulation, Phase Change and Microstructure},
  author = {Avrami, Melvin},
  year = {1941},
  volume = {9},
  pages = {177--184}
}

@article{babu2019Simulation,
  title = {Simulation of {{Ti-6Al-4V Additive Manufacturing Using Coupled Physically Based Flow Stress}} and {{Metallurgical Model}}},
  author = {Babu, Bijish and Lundb{\"a}ck, Andreas and Lindgren, Lars-Erik},
  year = {2019},
  month = nov,
  journal = {Materials},
  volume = {12},
  number = {23},
  pages = {3844},
  publisher = {MDPI AG},
  issn = {1996-1944},
  doi = {10.3390/ma12233844},
  urldate = {2025-01-14},
  copyright = {https://creativecommons.org/licenses/by/4.0/},
  langid = {english}
}

@article{barriobero-vila2017Inducing,
  title = {Inducing {{Stable}} {$\alpha$} + {$\beta$} {{Microstructures}} during {{Selective Laser Melting}} of {{Ti-6Al-4V Using Intensified Intrinsic Heat Treatments}}},
  author = {{Barriobero-Vila}, Pere and Gussone, Joachim and Haubrich, Jan and Sandl{\"o}bes, Stefanie and Da Silva, Julio Cesar and Cloetens, Peter and Schell, Norbert and Requena, Guillermo},
  year = {2017},
  month = mar,
  journal = {Materials},
  volume = {10},
  number = {3},
  pages = {268},
  publisher = {Multidisciplinary Digital Publishing Institute},
  issn = {1996-1944},
  doi = {10.3390/ma10030268},
  urldate = {2025-11-14},
  copyright = {http://creativecommons.org/licenses/by/3.0/},
  langid = {english}
}

@article{benzing2019Hot,
  title = {Hot Isostatic Pressing ({{HIP}}) to Achieve Isotropic Microstructure and Retain as-Built Strength in an Additive Manufacturing Titanium Alloy ({{Ti-6Al-4V}})},
  author = {Benzing, Jake and Hrabe, Nik and Quinn, Timothy and White, Ryan and Rentz, Ross and Ahlfors, Magnus},
  year = {2019},
  month = dec,
  journal = {Materials Letters},
  volume = {257},
  pages = {126690},
  issn = {0167-577X},
  doi = {10.1016/j.matlet.2019.126690},
  urldate = {2025-11-13}
}

@article{bierwischDevelopment,
  title = {Development and {{Validation}} of a {{One-Dimensional Finite Difference Simulation Scheme}} for {{Polymer Laser Powder Bed Fusion}} with {{Application}} to the {{Effect}} of the {{Inter Layer Time}}},
  author = {Bierwisch, Claas and Dietemann, Bastien and Gr{\"u}newald, Moritz and Schl{\"o}r, Christian and Rudloff, Johannes},
  journal = {Advanced Engineering Materials},
  volume = {n/a},
  number = {n/a},
  pages = {2401285},
  issn = {1527-2648},
  doi = {10.1002/adem.202401285},
  urldate = {2024-10-11},
  copyright = {{\copyright} 2024 The Author(s). Advanced Engineering Materials published by Wiley-VCH GmbH},
  langid = {english}
}

@article{brudler2024Systematic,
  title = {Systematic Investigation of Performance and Productivity in Laser Powder Bed Fusion of {{Ti6Al4V}} up to 300\,{\textmu}m Layer Thickness},
  author = {Brudler, S. and Medvedev, A. E. and Pandelidi, C. and Piegert, S. and Illston, T. and Qian, M. and Brandt, M.},
  year = {2024},
  month = sep,
  journal = {Journal of Materials Processing Technology},
  volume = {330},
  pages = {118450},
  issn = {0924-0136},
  doi = {10.1016/j.jmatprotec.2024.118450},
  urldate = {2025-11-13}
}

@article{charles2008modelling,
  title = {Modelling {{Ti-6Al-4V}} Microstructure by Evolution Laws Implemented as Finite Element Subroutines:: {{Application}} to {{TIG}} Metal Deposition},
  author = {Charles, Corinne and J{\"a}rvstr{\aa}t, Niklas},
  year = {2008}
}

@article{charlesmurgau2019Temperature,
  title = {Temperature and {{Microstructure Evolution}} in {{Gas Tungsten Arc Welding Wire Feed Additive Manufacturing}} of {{Ti-6Al-4V}}},
  author = {Charles Murgau, Corinne and Lundb{\"a}ck, Andreas and {\AA}kerfeldt, Pia and Pederson, Robert},
  year = {2019},
  month = oct,
  journal = {Materials},
  volume = {12},
  number = {21},
  pages = {3534},
  publisher = {MDPI AG},
  issn = {1996-1944},
  doi = {10.3390/ma12213534},
  urldate = {2025-01-14},
  copyright = {https://creativecommons.org/licenses/by/4.0/},
  langid = {english}
}

@article{chen2022Deciphering,
  title = {Deciphering the Transformation Pathway in Laser Powder-Bed Fusion Additive Manufacturing of {{Ti-6Al-4V}} Alloy},
  author = {Chen, J. and Fabijanic, D. and Zhang, T. and Lui, E. W. and Brandt, M. and Xu, W.},
  year = {2022},
  month = oct,
  journal = {Additive Manufacturing},
  volume = {58},
  pages = {103041},
  issn = {2214-8604},
  doi = {10.1016/j.addma.2022.103041},
  urldate = {2025-10-28}
}

@article{dhiman2024Microstructure,
  title = {Microstructure Control in Additively Manufactured {{Ti-6Al-4V}} during High-Power Laser Powder Bed Fusion},
  author = {Dhiman, Sahil and Chinthapenta, Viswanath and Brandt, Milan and Fabijanic, Daniel and Xu, Wei},
  year = {2024},
  month = sep,
  journal = {Additive Manufacturing},
  volume = {96},
  pages = {104573},
  issn = {2214-8604},
  doi = {10.1016/j.addma.2024.104573},
  urldate = {2025-03-12}
}

@article{esmaeilzadeh2023Insitu,
  title = {{\emph{In-Situ}} Selective Laser Heat Treatment for Microstructural Control of Additively Manufactured {{Ti-6Al-4V}}},
  author = {Esmaeilzadeh, Reza and {Hamidi-Nasab}, Milad and {de Formanoir}, Charlotte and Schlenger, Lucas and Van Petegem, Steven and Navarre, Claire and Cayron, Cyril and Casati, Nicola and Grolimund, Daniel and Log{\'e}, Roland E.},
  year = {2023},
  month = sep,
  journal = {Additive Manufacturing},
  volume = {78},
  pages = {103882},
  issn = {2214-8604},
  doi = {10.1016/j.addma.2023.103882},
  urldate = {2024-10-16}
}

@article{esmaeilzadehArchitected,
  title = {Toward {{Architected Microstructures Using Advanced Laser Beam Shaping}} in {{Laser Powder Bed Fusion}} of {{Ti-6Al-4V}}},
  author = {Esmaeilzadeh, Reza and Jhabvala, Jamasp and Schlenger, Lucas and {van der Meer}, Mathijs and Boillat, Eric and Cayron, Cyril and Jamili, Amir Mohammad and Xiao, Junfeng and Log{\'e}, Roland E.},
  journal = {Advanced Functional Materials},
  volume = {n/a},
  number = {n/a},
  pages = {2420427},
  issn = {1616-3028},
  doi = {10.1002/adfm.202420427},
  urldate = {2025-04-30},
  copyright = {{\copyright} 2025 The Author(s). Advanced Functional Materials published by Wiley-VCH GmbH},
  langid = {english}
}

@article{fan2005Effect,
  title = {Effect of Phase Transformations on Laser Forming of {{Ti}}--{{6Al}}--{{4V}} Alloy},
  author = {Fan, Y. and Cheng, P. and Yao, Y. L. and Yang, Z. and Egland, K.},
  year = {2005},
  month = jul,
  journal = {Journal of Applied Physics},
  volume = {98},
  number = {1},
  publisher = {AIP Publishing},
  issn = {0021-8979, 1089-7550},
  doi = {10.1063/1.1944202},
  urldate = {2025-01-14},
  langid = {english}
}

@misc{Figure,
  title = {Figure 4 {{Stress-strain}} Curve for the Wrought and {{SLM Ti-6Al-4V}}},
  journal = {ResearchGate},
  urldate = {2025-04-14},
  howpublished = {https://www.researchgate.net/figure/Stress-strain-curve-for-the-wrought-and-SLM-Ti-6Al-4V\_fig3\_281574099},
  langid = {english}
}

@article{haubrich2019Rolea,
  title = {The Role of Lattice Defects, Element Partitioning and Intrinsic Heat Effects on the Microstructure in Selective Laser Melted {{Ti-6Al-4V}}},
  author = {Haubrich, Jan and Gussone, Joachim and {Barriobero-Vila}, Pere and K{\"u}rnsteiner, Philipp and J{\"a}gle, Eric A. and Raabe, Dierk and Schell, Norbert and Requena, Guillermo},
  year = {2019},
  month = apr,
  journal = {Acta Materialia},
  volume = {167},
  pages = {136--148},
  issn = {1359-6454},
  doi = {10.1016/j.actamat.2019.01.039},
  urldate = {2025-11-13}
}

@article{johnson1939reaction,
  title = {Reaction Kinetics in Process of Nucleation and Growth},
  author = {Johnson, William A},
  year = {1939},
  journal = {Transactions of Transactions of the American Institute of Mining and Metallurgical Engineers},
  volume = {135},
  pages = {416--458}
}

@article{kavas2025Physicsaware,
  title = {Physics-Aware Feedforward Dwell Time Adjustment for Mitigating Distortion in Additively Manufactured Cantilevers},
  author = {Kavas, Bar{\i}{\c s} and Witte, Lars and Balta, Efe C. and Tucker, Michael R. and Afrasiabi, Mohamadreza and Bambach, Markus},
  year = {2025},
  month = jul,
  journal = {Progress in Additive Manufacturing},
  issn = {2363-9520},
  doi = {10.1007/s40964-025-01270-7},
  urldate = {2025-11-12},
  langid = {english}
}

@article{kavas2026Layer,
title = {Layer-to-layer closed-loop switched heating and cooling control of the laser powder bed fusion process},
journal = {Additive Manufacturing},
volume = {119},
pages = {105124},
year = {2026},
issn = {2214-8604},
doi = {https://doi.org/10.1016/j.addma.2026.105124},
url = {https://www.sciencedirect.com/science/article/pii/S2214860426000503},
author = {Barış Kavas and Efe C. Balta and Lars Witte and Michael R. Tucker and John Lygeros and Markus Bambach}
}

@article{kelly2004Thermala,
  title = {Thermal and {{Microstructure Modeling}} of {{Metal Deposition Processes}} with {{Application}} to {{Ti-6Al-4V}}},
  author = {Kelly, Shawn Michael},
  year = {2004},
  month = nov,
  publisher = {Virginia Tech},
  urldate = {2025-01-14},
  langid = {english}
}

@misc{khrenov2024Trajectory,
  title = {Trajectory {{Optimization}} for {{Spatial Microstructure Control}} in {{Electron Beam Metal Additive Manufacturing}}},
  author = {Khrenov, Mikhail and Tan, Moon and Fitzwater, Lauren and Hobdari, Michelle and Narra, Sneha Prabha},
  year = {2024},
  month = oct,
  number = {arXiv:2410.18207},
  eprint = {2410.18207},
  primaryclass = {eess},
  publisher = {arXiv},
  doi = {10.48550/arXiv.2410.18207},
  urldate = {2025-01-17},
  archiveprefix = {arXiv},
  langid = {english}
}

@inproceedings{klusemann2018Stability,
  title = {Stability of Phase Transformation Models for {{Ti-6Al-4V}} under Cyclic Thermal Loading Imposed during Laser Metal Deposition},
  booktitle = {{{Proceedings of the 21st International Esaform Conference on Material Forming}}: {{ESAFORM}} 2018},
  author = {Klusemann, Benjamin and Bambach, Markus},
  year = {2018},
  pages = {140012},
  address = {Palermo, Italy},
  doi = {10.1063/1.5035004},
  urldate = {2023-02-21},
  langid = {english}
}

@article{koistinen1959general,
  title = {A General Equation Prescribing Extend of Austenite-Martensite Transformation in Pure {{Fe-C}} Alloys and Plain Carbon Steels},
  author = {Koistinen, Donald P},
  year = {1959},
  journal = {Acta metallurgica},
  volume = {7},
  pages = {50--60}
}

@article{kolmogorov1937statistical,
  title = {On the Statistical Theory of the Crystallization of Metals},
  author = {Kolmogorov, Andrei Nikolaevich},
  year = {1937},
  journal = {Bull Acad Sci URSS (Cl Sci Math Nat)},
  volume = {3},
  pages = {335}
}

@article{lee2024Extreme,
  title = {Extreme Gradient Boosting-Based Multiscale Heat Source Modeling for Analysis of Solid-State Phase Transformation in Additive Manufacturing of {{Ti-6Al-4V}}},
  author = {Lee, Yeon Su and Lee, Kang-Hyun and Chung, Min Gyu and Yun, Gun Jin},
  year = {2024},
  month = mar,
  journal = {Journal of Manufacturing Processes},
  volume = {113},
  pages = {319--345},
  issn = {1526-6125},
  doi = {10.1016/j.jmapro.2024.01.044},
  urldate = {2025-01-10}
}

@article{leonor2024GOMELTa,
  title = {{{GO-MELT}}: {{GPU-optimized}} Multilevel Execution of {{LPBF}} Thermal Simulations},
  shorttitle = {{{GO-MELT}}},
  author = {Leonor, Joseph P. and Wagner, Gregory J.},
  year = {2024},
  month = jun,
  journal = {Computer Methods in Applied Mechanics and Engineering},
  volume = {426},
  pages = {116977},
  issn = {0045-7825},
  doi = {10.1016/j.cma.2024.116977},
  urldate = {2025-11-12}
}

@article{malinov2001Differential,
  title = {Differential Scanning Calorimetry Study and Computer Modeling of {$\beta$} {$\Rightarrow$} {$\alpha$} Phase Transformation in a {{Ti-6Al-4V}} Alloy},
  author = {Malinov, S. and Guo, Z. and Sha, W. and Wilson, A.},
  year = {2001},
  month = apr,
  journal = {Metallurgical and Materials Transactions A},
  volume = {32},
  number = {4},
  pages = {879--887},
  issn = {1543-1940},
  doi = {10.1007/s11661-001-0345-x},
  urldate = {2025-01-15},
  langid = {english}
}

@article{medvedev2025Interlayer,
  title = {Interlayer Time as a Robust, Geometry-Agnostic Predictor of Microstructural and Mechanical Properties Evolution in {{PBF-LB}}/{{M Ti6Al4V}} Alloy},
  author = {Medvedev, A. E. and Brudler, S. and Piegert, S. and Illston, T. and Qian, M. and Brandt, M.},
  year = {2025},
  month = jun,
  journal = {Journal of Materials Processing Technology},
  volume = {340},
  pages = {118858},
  issn = {0924-0136},
  doi = {10.1016/j.jmatprotec.2025.118858},
  urldate = {2025-04-17}
}

@article{mozaffar2023Differentiable,
  title = {Differentiable Simulation for Material Thermal Response Design in Additive Manufacturing Processes},
  author = {Mozaffar, Mojtaba and Liao, Shuheng and Jeong, Jihoon and Xue, Tianju and Cao, Jian},
  year = {2023},
  month = jan,
  journal = {Additive Manufacturing},
  volume = {61},
  pages = {103337},
  issn = {2214-8604},
  doi = {10.1016/j.addma.2022.103337},
  urldate = {2024-08-15}
}

@article{munk2022Geometry,
  title = {Geometry {{Effect}} on {{Microstructure}} and {{Mechanical Properties}} in {{Laser Powder Bed Fusion}} of {{Ti-6Al-4V}}},
  author = {Munk, Juri and Breitbarth, Eric and Siemer, Tobias and Pirch, Norbert and H{\"a}fner, Constantin},
  year = {2022},
  month = mar,
  journal = {Metals},
  volume = {12},
  number = {3},
  pages = {482},
  issn = {2075-4701},
  doi = {10.3390/met12030482},
  urldate = {2025-01-10},
  copyright = {https://creativecommons.org/licenses/by/4.0/},
  langid = {english}
}

@article{murgau2012Model,
  title = {A Model for {{Ti}}--{{6Al}}--{{4V}} Microstructure Evolution for Arbitrary Temperature Changes},
  author = {Murgau, C Charles and Pederson, R and Lindgren, L E},
  year = {2012},
  month = jul,
  journal = {Modelling and Simulation in Materials Science and Engineering},
  volume = {20},
  number = {5},
  pages = {055006},
  issn = {0965-0393, 1361-651X},
  doi = {10.1088/0965-0393/20/5/055006},
  urldate = {2023-02-21},
  langid = {english}
}

@article{nahr2023Geometrical,
  title = {Geometrical {{Influence}} on {{Material Properties}} for {{Ti6Al4V Parts}} in {{Powder Bed Fusion}}},
  author = {Nahr, Florian and Rasch, Michael and Burkhardt, Christian and Renner, Jakob and Baumg{\"a}rtner, Benjamin and Hausotte, Tino and K{\"o}rner, Carolin and Steinmann, Paul and Mergheim, Julia and Schmidt, Michael and Markl, Matthias},
  year = {2023},
  month = jun,
  journal = {Journal of Manufacturing and Materials Processing},
  volume = {7},
  number = {3},
  pages = {82},
  publisher = {Multidisciplinary Digital Publishing Institute},
  issn = {2504-4494},
  doi = {10.3390/jmmp7030082},
  urldate = {2025-01-10},
  copyright = {http://creativecommons.org/licenses/by/3.0/},
  langid = {english}
}

@article{nahr2025Advanced,
  title = {Advanced Process Control in Laser-Based Powder Bed Fusion--{{Smart Fusion}} Feedback-Loop Control as a Path to Uniform Properties for Complex Structures?},
  author = {Nahr, Florian and Novotny, Tobias and Kunz, Dominik and Kleinhans, Ulrich and Chechik, Lova and Bartels, Dominic and Schmidt, Michael},
  year = {2025},
  month = jan,
  journal = {Journal of Materials Research and Technology},
  volume = {34},
  pages = {604--618},
  issn = {2238-7854},
  doi = {10.1016/j.jmrt.2024.12.014},
  urldate = {2025-11-13}
}

@article{nitzler2021Novel,
  title = {A Novel Physics-Based and Data-Supported Microstructure Model for Part-Scale Simulation of Laser Powder Bed Fusion of {{Ti-6Al-4V}}},
  author = {Nitzler, Jonas and Meier, Christoph and M{\"u}ller, Kei W. and Wall, Wolfgang A. and Hodge, N. E.},
  year = {2021},
  month = jul,
  journal = {Advanced Modeling and Simulation in Engineering Sciences},
  volume = {8},
  number = {1},
  pages = {16},
  issn = {2213-7467},
  doi = {10.1186/s40323-021-00201-9},
  urldate = {2024-07-26},
  langid = {english}
}

@article{noll2024Thermodynamically,
  title = {A Thermodynamically Consistent Phase Transformation Model for Multiphase Alloys: Application to Ti-6Al-4V in Laser Powder Bed Fusion Processes},
  shorttitle = {A Thermodynamically Consistent Phase Transformation Model for Multiphase Alloys},
  author = {Noll, Isabelle and Bartel, Thorsten and Menzel, Andreas},
  year = {2024},
  month = dec,
  journal = {Computational Mechanics},
  volume = {74},
  number = {6},
  pages = {1319--1338},
  issn = {1432-0924},
  doi = {10.1007/s00466-024-02479-z},
  urldate = {2025-01-10},
  langid = {english}
}

@article{olleak2024Understandinga,
  title = {Understanding the Role of Geometry and Interlayer Cooling Time on Microstructure Variations in {{LPBF Ti6Al4V}} through Part-Scale Scan-Resolved Thermal Modeling},
  author = {Olleak, Alaa and Adcock, Evan and Hinnebusch, Shawn and Dugast, Florian and Rollett, Anthony D. and To, Albert C.},
  year = {2024},
  month = apr,
  journal = {Additive Manufacturing Letters},
  volume = {9},
  pages = {100197},
  issn = {2772-3690},
  doi = {10.1016/j.addlet.2024.100197},
  urldate = {2025-01-10}
}

@article{proell2024Highlya,
  title = {A Highly Efficient Computational Approach for Fast Scan-Resolved Microstructure Predictions in Metal Additive Manufacturing on the Scale of Real Parts},
  author = {Proell, Sebastian D. and Brotz, Julian and Kronbichler, Martin and Wall, Wolfgang A. and Meier, Christoph},
  year = {2024},
  month = jul,
  journal = {Additive Manufacturing},
  volume = {92},
  pages = {104380},
  issn = {2214-8604},
  doi = {10.1016/j.addma.2024.104380},
  urldate = {2025-01-14}
}

@article{promoppatum2022Understanding,
  title = {Understanding Size-Dependent Thermal, Microstructural, Mechanical Behaviors of Additively Manufactured {{Ti-6Al-4V}} from Experiments and Thermo-Metallurgical Simulation},
  author = {Promoppatum, Patcharapit and Taprachareon, Krisda and Chayasombat, Bralee and Tanprayoon, Dhritti},
  year = {2022},
  month = mar,
  journal = {Journal of Manufacturing Processes},
  volume = {75},
  pages = {1162--1174},
  issn = {1526-6125},
  doi = {10.1016/j.jmapro.2022.01.068},
  urldate = {2025-01-14}
}

@article{ransenigo2022meltpool,
  author  = {Ransenigo, Chiara and Tocci, Marialaura and Palo, Filippo and Ginestra, Paola and Ceretti, Elisabetta and Gelfi, Marcello and Pola, Annalisa},
  title   = {Evolution of Melt Pool and Porosity During Laser Powder Bed Fusion of Ti6Al4V Alloy: Numerical Modelling and Experimental Validation},
  journal = {Lasers in Manufacturing and Materials Processing},
  year    = {2022},
  volume  = {9},
  pages   = {481--502},
  doi     = {10.1007/s40516-022-00185-3}
}

@article{scheel2023Advancing,
  title = {Advancing Efficiency and Reliability in Thermal Analysis of Laser Powder-Bed Fusion},
  author = {Scheel, Pooriya and Wrobel, Rafal and Rheingans, Bastian and Mayer, Thomas and Leinenbach, Christian and Mazza, Edoardo and Hosseini, Ehsan},
  year = {2023},
  month = dec,
  journal = {International Journal of Mechanical Sciences},
  volume = {260},
  pages = {108583},
  issn = {0020-7403},
  doi = {10.1016/j.ijmecsci.2023.108583},
  urldate = {2023-10-23}
}

@article{vanini2025Local,
  title = {Local Microstructure Engineering of Super Duplex Stainless Steel via Dual Laser Powder Bed Fusion -- {{An}} Analytical Modeling and Experimental Approach},
  author = {Vanini, Michele and Searle, Samuel and Vanmunster, Lars and Vanmeensel, Kim and Vrancken, Bey},
  year = {2025},
  month = aug,
  journal = {Additive Manufacturing},
  volume = {112},
  pages = {104994},
  issn = {2214-8604},
  doi = {10.1016/j.addma.2025.104994},
  urldate = {2025-11-12}
}

@article{vastola2016Modelinga,
  title = {Modeling the {{Microstructure Evolution During Additive Manufacturing}} of {{Ti6Al4V}}: {{A Comparison Between Electron Beam Melting}} and {{Selective Laser Melting}}},
  shorttitle = {Modeling the {{Microstructure Evolution During Additive Manufacturing}} of {{Ti6Al4V}}},
  author = {Vastola, G. and Zhang, G. and Pei, Q. X. and Zhang, Y.-W.},
  year = {2016},
  month = may,
  journal = {JOM},
  volume = {68},
  number = {5},
  pages = {1370--1375},
  issn = {1543-1851},
  doi = {10.1007/s11837-016-1890-5},
  urldate = {2025-01-14},
  langid = {english}
}

@article{vrancken2012Heat,
  title = {Heat Treatment of {{Ti6Al4V}} Produced by {{Selective Laser Melting}}: {{Microstructure}} and Mechanical Properties},
  shorttitle = {Heat Treatment of {{Ti6Al4V}} Produced by {{Selective Laser Melting}}},
  author = {Vrancken, Bey and Thijs, Lore and Kruth, Jean-Pierre and Van Humbeeck, Jan},
  year = {2012},
  month = nov,
  journal = {Journal of Alloys and Compounds},
  volume = {541},
  pages = {177--185},
  issn = {0925-8388},
  doi = {10.1016/j.jallcom.2012.07.022},
  urldate = {2025-11-13}
}

@incollection{vrancken2016Preheating,
  title = {Preheating of {{Selective Laser Melted Ti6Al4V}}: {{Microstructure}} and {{Mechanical Properties}}},
  shorttitle = {Preheating of {{Selective Laser Melted Ti6Al4V}}},
  booktitle = {Proceedings of the 13th {{World Conference}} on {{Titanium}}},
  author = {Vrancken, Bey and Buls, Sam and Kruth, Jean-Pierre and Humbeeck, Jan Van},
  editor = {Venkatesh, Vasisht and Pilchak, Adam L. and Allison, John E. and Ankem, Sreeramamurthy and Boyer, Rodney and Christodoulou, Julie and Fraser, Hamish L. and Imam, M. Ashraf and Kosaka, Yoji and Rack, Henry J. and Chatterjee, Amit and Woodfield, Andy},
  year = {2016},
  month = may,
  edition = {1},
  pages = {1269--1277},
  publisher = {Wiley},
  doi = {10.1002/9781119296126.ch215},
  urldate = {2025-11-25},
  isbn = {978-1-119-28326-3 978-1-119-29612-6},
  langid = {english}
}

@article{wood2022Controllability,
  title = {On the {{Controllability}} and {{Observability}} of {{Temperature States}} in {{Metal Powder Bed Fusion}}},
  author = {Wood, Nathaniel and Hoelzle, David J.},
  year = {2022},
  month = dec,
  journal = {Journal of Dynamic Systems, Measurement, and Control},
  volume = {145},
  number = {031002},
  issn = {0022-0434},
  doi = {10.1115/1.4056326},
  urldate = {2025-04-16}
}

@article{xu2015Additivea,
  title = {Additive Manufacturing of Strong and Ductile {{Ti}}--{{6Al}}--{{4V}} by Selective Laser Melting via in Situ Martensite Decomposition},
  author = {Xu, W. and Brandt, M. and Sun, S. and Elambasseril, J. and Liu, Q. and Latham, K. and Xia, K. and Qian, M.},
  year = {2015},
  month = feb,
  journal = {Acta Materialia},
  volume = {85},
  pages = {74--84},
  issn = {1359-6454},
  doi = {10.1016/j.actamat.2014.11.028},
  urldate = {2025-11-13}
}

@article{xu2017Situ,
  title = {{\emph{In Situ}} Tailoring Microstructure in Additively Manufactured {{Ti-6Al-4V}} for Superior Mechanical Performance},
  author = {Xu, W. and Lui, E. W. and Pateras, A. and Qian, M. and Brandt, M.},
  year = {2017},
  month = feb,
  journal = {Acta Materialia},
  volume = {125},
  pages = {390--400},
  issn = {1359-6454},
  doi = {10.1016/j.actamat.2016.12.027},
  urldate = {2025-03-19}
}

@article{yang2021Processstructure,
  title = {Towards a Process-Structure Model for {{Ti-6Al-4V}} during Additive Manufacturing},
  author = {Yang, Xinyu and Barrett, Richard A. and Tong, Mingming and Harrison, Noel M. and Leen, Sean B.},
  year = {2021},
  month = jan,
  journal = {Journal of Manufacturing Processes},
  volume = {61},
  pages = {428--439},
  issn = {1526-6125},
  doi = {10.1016/j.jmapro.2020.11.033},
  urldate = {2025-01-14}
}

@article{zafari2019Controlling,
  title = {Controlling Martensitic Decomposition during Selective Laser Melting to Achieve Best Ductility in High Strength {{Ti-6Al-4V}}},
  author = {Zafari, A. and Barati, M. R. and Xia, K.},
  year = {2019},
  month = jan,
  journal = {Materials Science and Engineering: A},
  volume = {744},
  pages = {445--455},
  issn = {0921-5093},
  doi = {10.1016/j.msea.2018.12.047},
  urldate = {2025-11-13}
}


\end{document}